\documentclass{article}
\usepackage{graphicx}
\usepackage{cite}
\usepackage{hyperref}
\usepackage[T1]{fontenc}
\usepackage{amsfonts}
\newtheorem{theorem}{Theorem}
\newtheorem{lemma}{Lemma}
\newtheorem{remark}{Remark}
\newtheorem{corollary}{Corollary}
\usepackage{amsmath}
\author{Vadim Bobrovskiy\thanks{Department of Management, Economics and Industrial Engineering, Politecnico di Milano, Milan, 20156, Italy.} \thanks{Cosmetecor UK, London, England, W1H 1PJ.}
\and Juan Galvis\thanks{Departamento de Matem\'{a}ticas, Universidad Nacional de Colombia,  
Bogot\'{a} D.C.}
\and Alexey Kaplin\footnotemark[2] \and
Alexander Sinitsyn\footnotemark[3] \and Marco Tognoli\thanks{
Energy Engineering Department, Politecnico di Milano, Milan, 20156, Italy.} \and 
Paolo Trucco\footnotemark[1]
}

\title{Mathematical modelling of proton migration in Earth mantle} %

\begin{document}

%%%%%%%%%%%%%%%%%%%%%%%%%%%%%%%%%%%%%%%%%%%%
%                         THE TOP MATTER                         %
%%%%%%%%%%%%%%%%%%%%%%%%%%%%%%%%%%%%%%%%%%%%%%%%%%%%%%%%%%%%%%%%%%

%
%\runningtitle{Mathematical modelling of proton migration in Earth mantle}

\maketitle

\begin{abstract}
In the study, we address the mathematical problem of proton migration in the Earth’s mantle and suggest a prototype for exploring the Earth’s interior to map the effects of superionic proton conduction. The problem can be mathematically solved by deriving the self-consistent electromagnetic field potential U(x,t) and then reconstructing the distribution function f(x, v, t). Reducing the Vlasov-Maxwell system of equations to non-linear sh-Gordon hyperbolic and transport equations, the propagation of a non-linear wavefront within the domain and transport of the boundary conditions in the form of a non-linear wave are examined. By computing a 3D model and through Fourier-analysis, the spatial and electrical characteristics of potential U(x, t) are investigated. The numerical results are compared to the Fourier transformed quantities of the potential (V) obtained through field observations of the electric potential (Kuznetsov method). The non-stationary solutions for the forced oscillation of two-component system, and therefore, the oscillatory strengths of two types of charged particles can be usefully addressed by the proposed mathematical model. Moreover, the model, along with data analysis of the electric potential observations and probabilistic seismic hazard maps, can be used to develop an advanced seismic risk metric.
\end{abstract}

{{\bf Keywords}: Mathematical model, Vlasov-Maxwell equations, harmonic oscillator, time and 3D-space discretization, finite element space, non-linear hyperbolic sh-Gordon equation, transport equation, boundary-value problem, upper-lower solution}
%\subjclass{35L05, 35L70}

\maketitle
%%%%%%%%%%%%%%%%%%%%%%%%%%%%%%%%%%%%%%%%%%%%%%%%%%%%%%%%%%%%%%%%%%
% \section*{Introduction}                                        %
% \section{About the "head" of your paper}                       %
% \subsection{Your private macros (the preamble)}                %
% \subsection{The top matter}                                    %
% \section{About the "body" of your paper}                       %
% \subsection{The environments}                                  %
% \subsection{specific macros}                                   %
% \subsection{Cross reference and bibliography}                  %
% \subsection{Including postscript files}                        %
%%%%%%%%%%%%%%%%%%%%%%%%%%%%%%%%%%%%%%%%%%%%%%%%%%%%%%%%%%%%%%%%%%
%                              THE TEXT                          %
%%%%%%%%%%%%%%%%%%%%%%%%%%%%%%%%%%%%%%%%%%%%%%%%%%%%%%%%%%%%%%%%%%

\section*{Introduction}

Recent observations of prominent electrical conductivity due to superionic proton conduction in the host hydrous mineral (iron oxide-hydroxide ) \cite{Hou-Nature2021, He-Nature2022} in the deep Earth mantle have brought to concrete realization a paradigm and prediction by Vernadski 100 years ago \cite{vernadsky1912gas, larin1980hypothesis}. It is widely accepted that water in the mantle is stored in hydrous minerals. Recently, it has been reported that hydrous minerals can bring water into depths greater than 1250 km (lower mantle) \cite{Nishi-Nature2014}. In the deep Earth mantle, hydrogen is stored as hydroxyl ($OH^{-}$) in hydrous or other non-hydrous minerals. However, the amount of hydrogen in the Earth's deep interior and its form (molecular or atomic) are unclear. Hou et al. \cite{Hou-Nature2021} demonstrated that the protons in the FeOOH mineral structure moved freely through a lattice framework of oxygen and other cations. The free movement of hydrogen within water ice crystal has been recognized as superionic proton conduction \cite{Sugimura-Chem2012, Millot-Nature2018}. Furthermore, it has been experimentally confirmed that superionicity may occur in hydrous minerals in the Earth's deep interior. Here, the H atom constitutes the unit of mass and energy transport, and the migration of electrons induced by hydrogen is more profound. High-electrical conductivity offers a new perspective for using electromagnetic data to explore the Earth's interior \cite{Hou-Nature2021}.

For the last 100 years, dipole electric lines ("antenna crosses") were utilized for measuring the electromagnetic field at the surface, induced by the flow of telluric currents. Most academic research investigates the induction of electro-telluric anomalies \cite{Surkov} where a unit of the domain in the subsurface is modelled as an antenna under the influence of the Earth-ionosphere (or magnetosphere) resonator. The passive telluric antenna has a range of application, from geological prospecting \cite{Nabighian} to investigating the non-seismic forecasting of earthquakes \cite{Uyeda, Varostos}. It is general consensus that detection of the repetitive electric  signals originating from the Earth's interior  using electric line "crosses" is a highly complex task, and a global monitoring network is also not available.. 

This study examines the topic from two research perspectives: addressing the mathematical problem of proton migration in the Earth’s mantle and suggesting a prototype for exploring the Earth’s interior to map the effects of superionic proton conduction. 

The development of a mathematical model for proton migration in the superionic phases \cite{Hou-Nature2021, Gao2022, He-Nature2022} is highly challenging. Through literature survey including experimental results \cite{Hou-Nature2021}, we concluded that mathematical modelling in superionicity research must endeavour to capture the collective motion of charged particles (protons). The primary variable of interest is the self-consistent electromagnetic field potential created by the charges themselves. 

The collective motion of charged particles is a manifestation of phase boundary motion in the bulk of the host material. Phase transition cannot occur simultaneously throughout the entire domain. Therefore, the self-consistent electromagnetic field potential changes with time in the same direction in which the phase boundary moves through the material. This significantly affects electron transport as a function of the concentration \cite{Hanfland1993, Hemley1994, Hou-Nature2021}. 

Ensemble of charged particles in time-dependent domains exhibits oscillatory or wave behaviour \cite{Fortuin}. Key research and results focus on the properties of wave equations that enhance the understanding of practical motivating problems, such as the propagation of electromagnetic waves with a periodically moving boundary. On the theoretical side, hyperbolic partial differential equations (PDE) are mainly explored for solving the mixed initial-boundary value problem concerned with time evolution in cylindrical regions \cite{Lee}. The construction of analytical solutions  remains challenging for the classical, periodic wave because of the possible instability of the solutions of wave equations. In this case, the main theoretical focus is on spectral methods for studying the driven pattern formation during resonance. In the study, we extend the classical mixed initial-boundary value problem for second-order non-linear hyperbolic equations. The theoretical scope includes the development of a mathematical model of the propagation of a self-consistent electromagnetic field potential $U(x,t)$ in a time-dependent domain with moving walls, where self-consistent electromagnetic field potential $U(x,t)$ is created by the charges of two types ($H^{+}$ and $OH^{-}$).  

\begin{figure}
\center{
\includegraphics[height=0.3\textwidth]{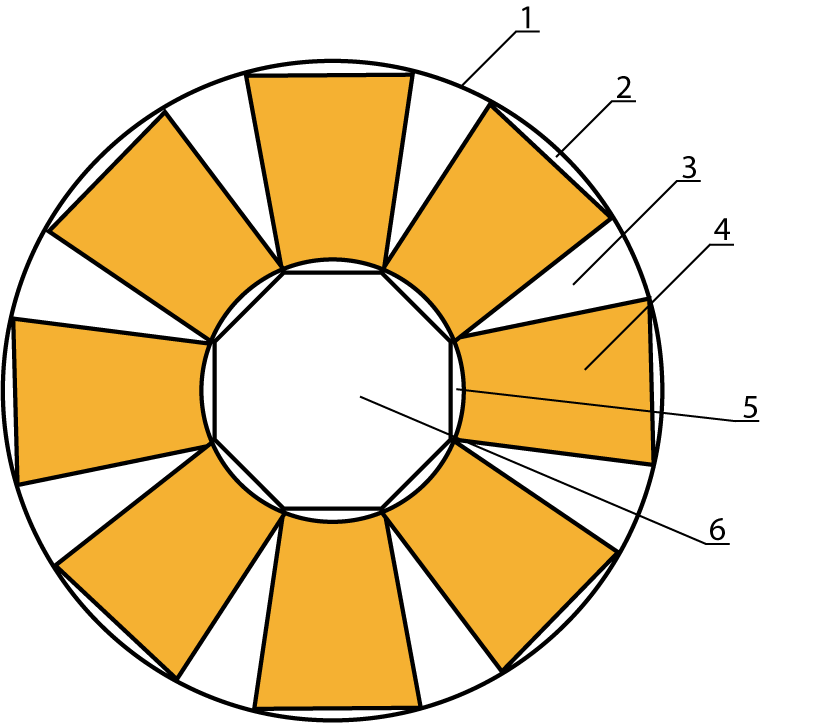}\quad
\includegraphics[height=0.3\textwidth]{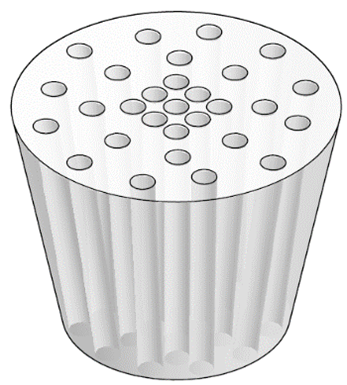}
}
\center{a)\phantom{xxxxxxxxxxxxxxxxxxxxx}b)}
\caption{a) Model of the Earth's interior, including 1. Surface of the Earth’s outer shell, 2. Space beneath the surface, 3. Space between the cones, 4. Model of the potential $U(x, t)$ dynamics (truncated cone domain), 5. Model of the current source (space between the inner shell (core) and bottom of the cone), and 6. Inner shell (core) of the Earth. b) Cylindrical channel geometry of the truncated cone domain under investigation.}
\label{figmodel-general}
\end{figure}

We refer to the evidence-based description of physical phenomena and Vlasov's theory \cite{vlasov1961many} for developing the mathematical model of the self-consistent electromagnetic field potential $U(x, t)$ induced in a truncated cone domain in the Earth's mantle through superionic proton conduction. The source of the collective motion of charged particles is localized at the core-mantle boundary (see figure \ref{figmodel-general}). The mathematical description addresses the following question: What happens when the potential is treated as a particle distribution function of two types of charges in a phase space where both the position and velocity approach unstable stationary states? This amounts to the study of a solution for Vlasov-Maxwell (VM) equations concerned with the de-stabilization of a homogeneous stationary state and the non-linear development of the wave governed by resonances, in general. The rationale is that these distribution functions should be i)relevant to two types of charged particles in a many-particle system, ii) governed by collision-less physics where the dominant physical interactions are caused by energy-induced Coulomb force for electrostatic interaction in an ensemble of charged particles, and iii) could provide basic building blocks to describe the quantitative properties of potential $U(x, t)$ and its complex dynamics. 

In the next section, we formulate the problem statement based on experimental results and a survey of the data in literature. Based on the problem statement, we treat the collective motion of charged particles on a mesoscopic level with distribution functions in the form of the VM system of equations. Time-dependence of the spatial domain implies that the problem is distributed in space and is therefore, infinite-dimensional. We supplement the VM system of equations with initial data and boundary conditions. 

For a special ansatz of the distribution functions involving the self-consistent electromagnetic field potential, we reduce the VM system of equations to non-linear sh-Gordon hyperbolic and transport equations. We solve this overdetermined system using the method of characteristics. From these results, we obtain the solution in the form of a travelling wave.  Further, we solve the initial-boundary value problem for a non-linear hyperbolic equation using the lower-upper solutions method under the condition that the self-consistent electromagnetic field potential $U(x, t)$ is in the form of a travelling wave. 

Using the sh-Gordon equation derived from the 3D wave equation, we numerically simulate a two-component system. We verify through computer simulations, using a complex 3D model, that the self-consistent electromagnetic field potential propagates within the spatial domain as a non-linear traveling wave, and that the boundary conditions move in time as a non-linear wave. Depending on the initial data selection, the non-linear wave takes the form of either nested hierarchies on a wavefront or waves mixing with resonant frequencies. The obtained numerical results are then compared with field observations (Kuznetsov method) \cite{kuznetsov1991practice} through Fourier transform. The Kuznetsov method, which involves field observation of the electric potential, has been investigated by the global station network (Cosmetecor) in Kamchatka, Italy, Altai, Fiji, and Sakhalin \cite{bobrovskiy2017nonstationary, Bobrovskiy2016thesis, Bobrovskiy2016book} over last 30 years. Non-stationary changes of electric potential and their relation to the major seismic events that accompany subduction of the Pacific plate under Kamchatka are taken as an illustrative example. Incorporating the proposed PDE model and statistical models, future exploration of realistic phenomena in Earth science becomes feasible. With the proposed mathematical (PDE) model, experimental results on the phase transition (charge-transfer state) in dense hydrogen obtained 30 years ago are discussed \cite{Hanfland1993, Hemley1994}. 

In Section 3, the VM system is reduced to non-linear hyperbolic and transport equations. In Section 4, the boundary value problem is investigated. Section 5 reports and discusses the non-linear numerical simulations, and the concluding remarks are finally presented.

\section{Problem statement}

\begin{figure}
\center{
\includegraphics[height=0.35\textwidth]{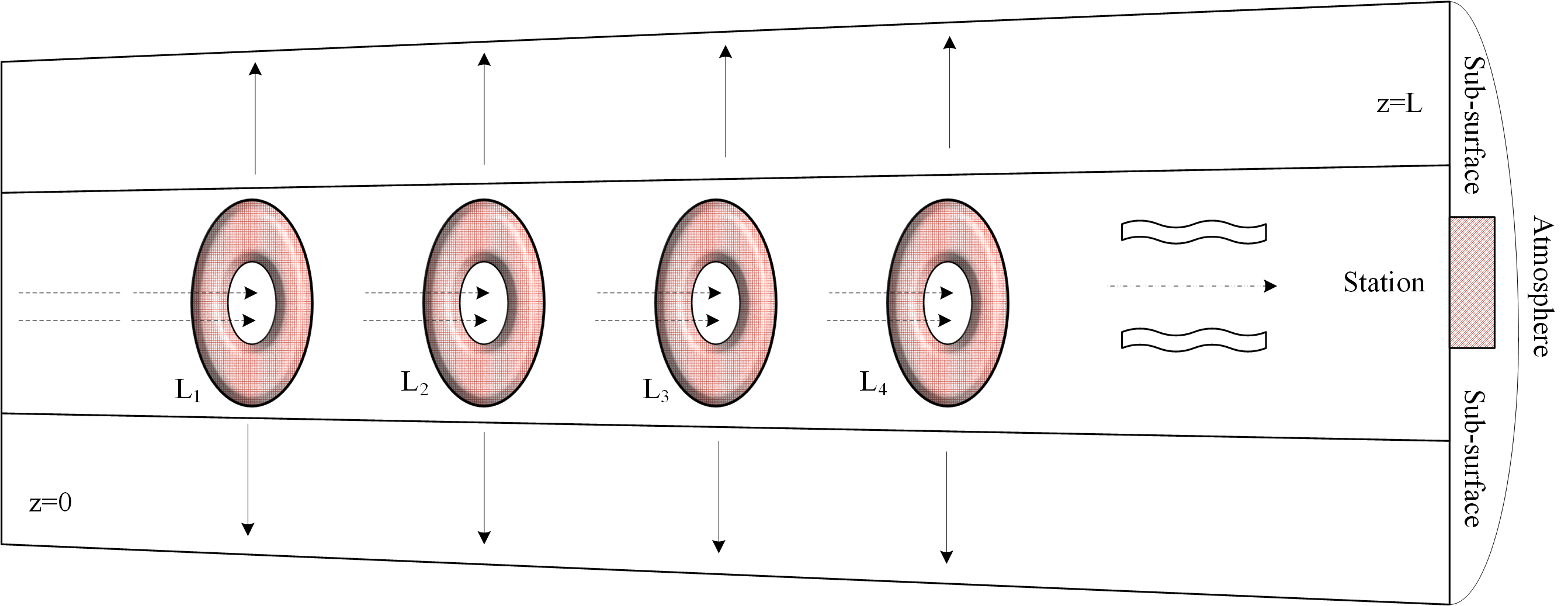}\quad}
\caption{Illustration of the problem statement where domain $\Omega$ is limited to the space between $L_{1}$ and $L_{4}$ (upper-mantle). The source of collective motion of the charged particles is localized at z=0 (core-mantle boundary). The double-dotted arrows denote vector $d$ whose direction coincides with the Z-axis of symmetry of the ansatz distribution function of the velocity and is positive vertically upwards. 
The directions of the time-dependent electric field are also denoted by double-dotted arrows in the plot. The solid-arrows represent the time-dependent self-consistent magnetic field $B(x, t)$. The single dotted-line and the waves represent the physical processes occurring in the Earth's asthenosphere and crust. The rectangle denotes the location of the multi-electrode station in the Earth's surface boundary. }
\label{figmodel-problemst}
\end{figure}

Let $\Omega$ be an anisotropic and multi-layered domain in the Earth's lower- and upper-mantle (see figure \ref{figmodel-problemst}), where $\Omega = L_{1} \cup L_{2} \cup L_{3} \cup L_{4}$. 
Under certain temperature and pressure, it comprises a fixed ratio of atoms in a defined spatial homogeneous arrangement. During a short period, the arrangement can be approximated to thread-like and plate-like crystalline structures. The arrangement migrates in a stepped manner from $L_{1}$ to $L_{4}$ sequentially. The quantitative determination of the full distribution of the “atomic-level anisotropy” in domain $\Omega$ is beyond the scope of this study.

Based on experimental results, we consider the positive ions as proton ($H^{+}$, atomic) condensates, analogous to electron condensates, and the negative ions as $OH^{-}$ (molecular) condensates. We introduce a distribution function $f_{i}$ that describes the behaviour of the proton and molecular condensates interacting through the Coulomb force, ignoring collisions. In the Vlasov theory \cite{vlasov1961many}, the distribution functions evolve in time as an entity and are not constructed from particle statistics. The main advantage is that the distribution function becomes noiseless. Therefore, distribution function $f_{i}(x, v, t)$ characterizes domain $\Omega$. 

Consider the self-consistent electromagnetic field potential $U(x, t)$ induced in domain $\Omega$ through superionic proton conduction. The source of collective motion of the charged particles is localized at the core-mantle boundary (z=0, see figure \ref{figmodel-problemst}). The superionic phase is not constant in time and is impulsive by nature. Therefore, self-consistent electromagnetic field potential is changed in time propagating in time-dependent domain with moving walls,  where self-consistent electromagnetic field potential is created by the charges of two types ($H^{+}$ and $OH^{-}$).

The self-consistent electromagnetic field, $E(x, t)$ and $B(x, t)$, is time dependent on the spatial boundary; the self-consistent electric field is aligned with the unit outward normal vector, the self-consistent magnetic field is orthogonal to the unit outward normal vector, and the distribution functions satisfy the reflection conditions with respect to the particle velocity. 

We will derive the self-consistent electromagnetic field potential $U(x,t)$ in domain $\Omega$ with the boundary condition on the boundary of domain $\partial\Omega$ and then we will reconstruct the distribution functions $f_{i}(x, v, t)$. 

We treat distribution function $f_{i}(x, v, t)$ only for one selected truncated cone in Figure \ref{figmodel-problemst} because all the truncated cones are symmetrical about the centre, which is the Earth's core (see Figure \ref{figmodel-general}). We consider a sub-vertical cylindrical truncated cone to extend from z=0 (core-mantle boundary) to z=L, i.e., approximately 70 km. Figure \ref{figmodel-problemst} shows the physical processes manifesting in the uppermost part of the cone (in the Earth's asthenosphere and the crust) as a single dotted line and waves, addressable by observing the electric potential. The multi-electrode Cosmetecor station (Kuznetsov method) records the electric potential in the sub-surface of the Earth's surface boundary.

\section{Vlasov-Maxwell system: Reduction to non-linear hyperbolic and transport equations}

We consider the classical solutions of the VM equations in $\Omega\times \mathbb{R}^3\times (0, T)$ with boundary conditions on the electromagnetic field on $\partial \Omega\times (0,T)$, where $\Omega=G\times {\mathbb R}$ is an infinite cylinder and $G\subset {\mathbb R}^{2}$ is a bounded domain with boundary $\partial G\in C^{1}$. 

\begin{equation}
\partial_{t}f_{i} + v\cdot \nabla_{x}f_{i} + \frac{q_{i}}{m_{i}}\left ( E + \frac{1}{c}v\times B \right )\cdot \nabla_{v}f_{i}= 0, \;\;\;i=1, 2,
\end{equation}
$$
(x\in \Omega, \;v\in {\mathbb R}^{3}, \; 0<t<T)
$$
\begin{equation}
\partial_{t}E= c\, {\rm rot} B - j,            \hspace{10.6cm}
\end{equation}
\begin{equation}
{\rm div} E = \rho,                            \hspace{12.0cm}
\end{equation}
\begin{equation}
\partial_{t}B=-c\;{\rm rot} E,                 \hspace{10.8cm}
\end{equation} 
\begin{equation}
{\rm div} B= 0,                                 \hspace{12.0cm}
\end{equation}
with initial conditions
\begin{equation}
f_{i}(x, v, t)|_{t=0}= f_{i}^{0}(x,v),\; E(x, t)|_{t=0}=E^{0}(x), \;  B(x,t)|_{t=0}=B^{0}(x) \;\;\;\; (x\in \bar{\Omega}, v\in {\mathbb R}^{3}).
\end{equation}

Here, $f_{i}\stackrel{\triangle}{=} f_{i}(x, v, t)$ is a distribution function; $x \stackrel {\triangle}{=}(x_{1}, x_{2}, x_{3}) \in \Omega$, $v\stackrel {\triangle}{=}(v_{x_{1}}, v_{x_{2}}, v_{x_{3}})\in {\mathbb R}^{3}$ are the position and velocity, respectively, of a particle; $E \stackrel\triangle{=} E(x, t)$ and $B \stackrel{\triangle}{=} B(x, t)$ are self-consistent electric and magnetic fields, respectively; $q_{1}>0$ and $m_{1}$ are the charge and mass, respectively, of a positive proton; $q_{2}<0$ and $m_{2}$ are the charge and mass, respectively, of a negative ion. 

The charge and current densities are defined as follows: 
\begin{equation}
\rho(x, t) = 4\pi\sum_{i=1}^{2}q_{i}\int_{{\mathbb R}^{3}}f_{i}dv, \;\;\;\;j(x, t)= 4\pi \sum_{i=1}^{2}\int_{{\mathbb R}^{3}}f_{i}vdv.
\end{equation}
   
We search for the distribution functions of protons and negative ions in the form
\begin{equation}
f_{i}=f_{i}(-\alpha_{i}|v|^{2}+ vd_{i}+ l_{i}U(x,t)), \;\;\; d_{i}\in {\mathbb R}^{3}, \;\;\; \alpha_{i}\in[0,\infty), \;\;\;l_{i}=\frac{\alpha_{i}mq_{i}}{\alpha m_{i}q}                    
\end{equation}
and the corresponding fields $E(x, t), B(x, t)$ satisfying (2)--(5). Function $U(x,t): [0, \infty[\times \Omega\rightarrow {\mathbb R}$ is an electromagnetic field potential that has to be determined. Constant vector $d_{i}$ defines the symmetry of the problem in the domain. 
We reduce the VM system for the ansatz of distribution function (8) to a hyperbolic equation  

\begin{equation}
\frac{\partial^{2}U}{\partial t^{2}}= c^{2}\triangle U+ \frac{2\pi q}{\alpha m}(d^{2}- 4\alpha^{2}c^{2}) \sum_{i=1}^{2}q_{i}\int_{\mathbb R^{3}} f_{i}dv,                                  
\end{equation}

and transport equation 

\begin{equation}
2\alpha\frac{\partial U}{\partial t}+(\nabla U, d)=0.                             
\label{eq11-9-9.8}
\end{equation}

Based on (9) and (10), we formulate the following theorem:
\vspace{0.2cm}

{\bf Theorem 1.}{\it Let $f_{i}(s)$ be arbitrary differential functions; moreover, 
$$
\int_{\mathbb R^{3}} f_{i}(-|v|^{2}+T) dv<\infty, \;\;\; T\in(-\infty, +\infty), \;\;\; \alpha_{i}\in[0,\infty), \;\;\; d_{i}\in {\mathbb R^{3}},
$$
$$
\alpha_{i}d= \alpha d_{i}, \;\;\; \alpha\stackrel{\triangle}{=}\alpha_{1}, \;\;\; d\stackrel{\triangle}{=}d_{1}.
$$
Then, every solution $U(x, t)$ of hyperbolic equation (9) with condition (10) corresponds to a solution of the system (1)--(5) in the form
\begin{equation}
f_{i}= f_{i}(-\alpha_{i}|v|^{2} +vd_{i}+ \lambda_{i}+ l_{i}U (x, t)),         \hspace{3.3cm}               \label{eq23-9-9.8}
\end{equation}
$$
\hspace{2.3cm} B=\frac{\gamma}{d^{2}}d +\frac{mc}{q(d^{2}-4\alpha^{2}c^{2})} [\nabla (U+\varphi_{0}(x))\times d], \hspace{4.3cm}
$$
\begin{equation}
 E=\frac{m}{2\alpha q(4\alpha^{2}c^{2}- d^{2})}\{\nabla(4\alpha^{2} c^{2}U+ d^{2}\varphi_{0}(r) - (\nabla U, d)d) \},                    
\label{eq24-9-9.8}
\end{equation}
where $\varphi_{0}(x)$ is an arbitrary function satisfying $\triangle \varphi_{0}= 0, \; \nabla\varphi_{0}\perp d$.}

The proof of the theorem is given in \cite{rudykh1989nonstationary}.

\begin{corollary}
{\it In a stationary case, (9) is transformed to the form 
\begin{equation}
\triangle U(x)= \frac{2\pi q}{\alpha mc^{2}}(4\alpha^{2}c^{2}- d^{2}) \sum_{i=1}^{2}q_{i}\int_{\mathbb R^{3}} f_{i}dv           
\end{equation}
with condition} 
\begin{equation}
(\nabla U, d)=0.                          
\end{equation}
\end{corollary}

\begin{remark}
 If 
$$
f_{i}=e^{s}, \;\;\; S=-\alpha_{i}|v|^{2} + vd_{i} +\lambda_{i}+ l_{i}U, \;\;\; l_{i}=\frac{\alpha_{i} m q_{i}}{\alpha m_{i}q},
$$
then, 
$$
\int_{\mathbb R^{3}}f_{i}dv =\left (\frac{\pi}{\alpha_{i}}\right )^{3/2}\exp\{d_{i}^{2}/4\alpha_{i} +\lambda_{i} +l_{i}U\}.
$$
\label{rm5_9.5}
\end{remark}
In this case, (9) can be expressed as follows: 
$$
\frac{\partial^{2}U}{\partial t^{2}} = c^{2}\triangle U+\frac{2\pi q}{\alpha m}(d^{2}- 4\alpha^{2}c^{2})\pi^{3/2}\sum_{i=1}^{2} q_{i}(\alpha_{i})^{-3/2}\exp\{d_{i}^{2}/4\alpha_{i}+ \lambda_{i}+l_{i}U\}.
$$
 For case $N=2$ (2D case), this equation is transformed to
\begin{equation}
\frac{\partial^{2}U}{\partial t^{2}} = c^{2} \triangle U + \lambda b(e^{U}- e^{lU}), \;\;\; l\in R^{-}, \;\;\;\lambda\in R^{+},                
\label{eq27-9-9.8}
\end{equation}
$$
b=\frac{2\pi q^{2}}{\alpha m}\left (\frac{\pi}{\alpha}\right )^{3/2}(d^{2}- 4\alpha^{2}c^{2})e^{d^{2}/4\alpha}.
$$
When $l=-1$, (\ref{eq27-9-9.8}) becomes an {\rm sh}-Gordon equation:
\begin{equation}
\frac{\partial^{2}U}{\partial t^{2}}=c^{2}\triangle U + 2\lambda b\sin hU.             
\label{eq28-9-9.8}
\end{equation}

\begin{remark}
In accordance with the conditions of Theorem 1, the scalar $\Phi$ and vector $A$ potentials are defined as follows:  
$$
\Phi=\frac{m}{2\alpha q(d^{2}- 2\alpha^{2}c^{2})}\{4\alpha^{2}c^{2}U(x,t)+ d^{2}\varphi_{0}\},
$$
\begin{equation}
A= \frac{mc}{q(d^{2}- 4\alpha^{2}c^{2})} d\{U(x,t)+\varphi_{0}\}+ \triangle\Theta(x),                                
\end{equation}
where 
\begin{equation}
\triangle\Theta(x)= \frac{\gamma}{d^{2}} (d_{2}x_{z}, d_{3}x_{x}, d_{1}x_{y})' +\nabla p(x), \;\;\; d\stackrel{\triangle}{=} (d_{1}, d_{2}, d_{3})                      
\end{equation}
and $p(x)$ is an arbitrary harmonic function. As function $U(x,t)$ satisfies (9), potentials $\Phi$ and $A$ are connected by Lorentz calibration 
$$
\frac{1}{c}\frac{\partial \Phi}{\partial t}+ {\rm div} A=0.
$$
\end{remark}
For analysing (16), we direct a constant vector $d\in {\mathbb R^{3}}$ along the $Z$ axis; i.e., we assume that $d\stackrel{\triangle}{=}(0, 0, d_{z})$. Moreover, solution $U(x, y, z, t)$ for (13) has the form
\begin{equation}
U=U(x, y, z-\frac{d}{2\alpha}t).                                     
\end{equation}
Solution (19) describes the wave spreading velocity in the positive direction along the $Z$ axis at constant velocity $d/2\alpha$ (direction of symmetry of domain $\Omega$), where $d/2\alpha<c$. Substituting $\xi=z-(d/2\alpha)t$, we reduce (16) to 
\begin{equation}
\frac{\partial^{2}U}{\partial x^{2}} +\frac{\partial^{2}U}{\partial y^{2}} +\frac{(4\alpha^{2}c^{2}- d^{2})}{4\alpha^{2}c^{2}} \frac{\partial^{2}U}{\partial\xi^{2}}=2\lambda p\sinh\; U,                      
\end{equation}
where   
$$
p\stackrel{\triangle}{=}\frac{2\pi q^{2}}{\alpha mc^{2}}\left (\frac{\pi}{\alpha}\right )^{3/2} (4\alpha^{2}c^{2}- d^{2})\exp(d^{2}/4\alpha)>0; \;\;\;\lambda\in R^{+}.
$$
Moreover, introducing a new variable $\eta=(4\alpha^{2}c^{2}/(4\alpha^{2}c^{2}-d^{2}))^{1/2}\xi$, we transform  (20)
\begin{equation}
\frac{\partial^{2}U}{\partial x^{2}}+ \frac{\partial^{2}U}{\partial y^{2}}+ \frac{\partial U}{\partial\eta^{2}}=2\lambda p\sinh \;U, \;\;\; U\stackrel{\triangle}{=} U(x, y, \eta).                   
\label{eq33-9-9.8}
\end{equation}
\vspace{0.5cm}

\section{Boundary value problem}
Here, we consider the classical solutions $(f_{1}, \ldots, f_{n}, E, B)$ of the VM system in a special form ((11), (12)), which we express as follows: 
\begin{equation}
f_{i}(x,v,t) =\hat{f}_{i}(-\alpha_{i}v^{2} +vd_{i}+l_{i}U(x,t)),            \hspace{3.2cm}                           
\end{equation}
\begin{equation}
E(x, t)= \frac{m}{2\alpha q(4\alpha^{2}c^{2}- d^{2})} \left (4\alpha^{2}c^{2}\nabla U(x, t)+ \partial_{t} U(x, t)d\right ),                               
\end{equation}
\begin{equation}
B(x, t)= \frac{\gamma}{d^{2}}d - \frac{mc}{q(4\alpha^{2}c^{2}- d^{2})}\nabla U(x, t)\times d, \hspace{3.4cm}              
\end{equation}
where functions $\hat{f}_{i}:\; {\mathbb R}\rightarrow [0,\infty[$ and vector $d\in {\mathbb R^{3}}\backslash \{0\}$ are given, and function $U:\; [0,\infty[\times \bar{\Omega}\rightarrow {\mathbb R}$ has to be defined. Assuming $\partial\Omega\in C^{1}$, we introduce the boundary conditions for the electromagnetic field
\begin{equation}
E(x,t)\times n_{\Omega}(x)=0, \;\;\; B(x,t) n_{\Omega}(x)=0, \;\;\;t\ge 0, \;\;\;x\in \partial\Omega,                                       
\end{equation}
and the specular reflection condition for the distribution function on the boundary 
\begin{equation}
f_{i}(t,x,v)= f_{i}(t, x, v-2(v n_{\Omega}(x))n_{\Omega}(x)), \;\;\;t\ge 0, \;\;\;x\in\partial\Omega, \;\;\;v\in {\mathbb R}^{3},                               
\end{equation}
where $n_{\Omega}$ is a unit vector normal to $\partial\Omega$.

{\bf Definition 1.} Vector function $\{E, B, f_{i}\}$ with
$$
E(x, t)\in C^2(\bar{\Omega}, [0, T]), \;\;\; B(x, t)\in C^2(\bar{\Omega}, [0, T]), \;\;\; f_{i}(x, v, t) \in C^1(\bar{\Omega}\times \mathbb{R}^3\times [0, T])
$$ 
is a classical solution for problems (1)--(6), (25), and (26), if $\{E, B, f_{i}\}$ satisfies VM equations (1)--(5), initial conditions (6), and boundary conditions (25)--(26). Moreover, vector function $\{U, f_{i}\}$ 
with 
$$
U(x,t)\in C^2(\bar{\Omega}, [0, T]), \;\;\; f_{i}(x, v, t) \in C^1(\bar{\Omega}\times {\mathbb R}^3\times [0, T])
$$
satisfies the boundary value problem (29), and $E, B, f_{i}$ are expressed in terms of $U(x,t)$ in (22)--(24). Adams and Fournier \cite{Adams} introduced definitions and theorems for Sobolev spaces and other related spaces of the function, which are reflected in our study. 

To prove the existence of classical solutions for (1)--(5) and (25)--(26), we apply the lower-upper solutions method developed for non-linear elliptic systems. In contrast to the stationary problem, the non-stationary one is more complicated because we need to add an equation of the first order (10) to non-linear wave equation (9). Hence, the problem is not ''strongly'' elliptic, and further development of the method of lower-upper solutions is needed.

\begin{lemma} 
{\it Let $\Omega\subset {\mathbb R}^{n}$ be a bounded domain with boundary $\partial\Omega\in C^{2,\alpha}, \; \alpha\in]0,1[$. Let $u_{0}\in C^{2,\alpha}(\bar{\Omega})$ and $h\in C_{loc}^{0,1}(\bar{\Omega}\times {\mathbb R})$ such that $h(x,\cdot)$ is a monotonically increasing function for every $x\in \Omega$. Then, boundary value problem 
\begin{equation}
\triangle u= h(\cdot, u(\cdot))\;\;\; {\rm in} \;\;\;\Omega                                
\label{eq39-9-9.8}
\end{equation}
$$
u=u_{0}\;\;\;{\rm on}\;\;\;\partial\Omega           
$$
possesses a unique solution} $u\in C^{2.\alpha}(\bar{\Omega})$.
\label{lm12_9.12}
\end{lemma}
\vspace{0.1cm}

{\it Proof.} Due to the monotonicity of $h$, it is easy to check whether there exist $p_{1}, p_{2}\in C^{0,\alpha}(\Omega)$ such that $p_{2}(x)\le 0\le p_{1}(x)$ and 
$$
h(x,s)\left \{\begin{array}{c}
\le p_{1}(x)\;\;{\rm for} \;\; s\le 0,\\[0.2cm]
\ge p_{2}(x)\;\; {\rm for} \;\; s\ge 0
\end{array}\right.
$$
for all $x\in \bar{\Omega}$. Let $u_{01}=\min(u_{0}, 0)$ and $u_{02}=\max(u_{0},0)$. Let $u_{k}\in C^{2,\alpha}(\bar{\Omega})$ be a solution of the linear boundary value problem for $k\in (1, 2)$
$$\left \{\begin{array}{c}
\triangle u_{k}= p_{k}\;\;{\rm in} \;\;\Omega, \\[0.2cm]
u_{k}=u_{0k}\;\;{\rm on} \;\;\partial\Omega.
\end{array}\right.
$$
According to the maximum principle, $u_{1}\le 0\le u_{2}$ in $\bar{\Omega}$. From the last one, it follows that $u_{1}$ is a lower solution and $u_{2}$ is an upper solution for (27). Then, from the theorem of existence (see \cite{Pao}, Theorem 7.1), it follows that (27) has a unique solution $u\in C^{2,\alpha}(\bar{\Omega})$. \hspace{3.8cm}                          $\Box$ 

\begin{remark}
Lemma 1 is a well-known statement and does not require additional comments. We only remark that the monotonicity condition for function $h(x,\cdot)$ for the VP system is first applied by \cite{sinitsyn2011kinetic}.
\end{remark}

We introduce the following conditions for function $\hat{f}: \; {\mathbb R}\rightarrow [0,\infty[:$

${\bf (f1)}$ $\; \hat{f}\in C^{1}({\mathbb R})$;

${\bf (f2)}$ $\;\forall u\in {\mathbb R}: \; f\in L^{1}(u,\infty)$;

${\bf (f3)}$ $\;f$ is a measurable function and $f(s)\le C e^{-s}$ for a.e. $s\in {\mathbb R}$;

${\bf (f4)}$ $\; f$ is decreasing, $f(0)=0$ and $\exists\mu\ge 0:\; \forall s\le 0:\; f(s)\le C|s|^{\mu}$.

\begin{lemma} 
(\cite{braasch1997semilineare}). {\it Let function $f: \; {\mathbb R}\rightarrow [0,\infty[$ and 
$$
h_{f}(u)=c\int_{\mathbb R^{3}}f(v^{2}+ vd+ u), \;\;\; u\in {\mathbb R}.
$$
Then, the following claims hold:

1. Assume conditions ${\bf (f2)}$ and ${\bf (f3)}$. Then, $h_{f}: \; {\mathbb R}\rightarrow {\mathbb R}$ is continuous and non-negative
\vspace{0.1cm}

$$
h_{f}(u)=\frac{c_{1}}{|d|} \int_{1}^{\infty}\int_{-|d|s^{2}}^{|d|s^{2}}s f(s+t+u) dt ds
$$
for all $u\in {\mathbb R}$.

2. Assume condition ${\bf (f3)}$. Let $\psi:\;{\mathbb R} \rightarrow [0,\infty[$ be a measurable function and $\psi\le f$ (a.e.). Then, $h_{\psi}\le h_{f}$.

3. Assume conditions ${\bf (f4)}$ and $|d|<1$. Then, the following conditions ${\bf (f2), (f3)}$, $h_{f}$ are continuously differentiable and $h_{f}(u)\le Ce^{-u}$ for all $u\in {\mathbb R}$ are satisfied.

4. Assume ${\bf (f4)}$ and $|d|<1$. Then, from ${(f4)}$, it follows that $h_{f}$ is a decreasing function and 
$$
|h_{f}(u)|\le C|u|^{\mu}
$$
for all $u\in {\mathbb R}$, where} $C=C(\mu, |d|)$.
\end{lemma}

\begin{lemma} 
{\it Let $\Omega\in {\mathbb R}^{2}$ with a smooth boundary $\partial\Omega \in C^{1}$. Let $\hat{f}_{1}, \ldots, \hat{f}_{n}: \; {\mathbb R}\rightarrow [0,\infty[$ satisfy conditions ${\bf (f1)-(f3)}$
and $|d|<1$. $h_{f}:\;\bar{\Omega}\times {\mathbb R}\rightarrow {\mathbb R}$ is given by 
\begin{equation}
h_{f}(x, U) =- \frac{2\pi q}{\alpha m}(4\alpha^{2}c^{2}- d^{2}) \sum_{i=1}^{2}q_{i}\int_{\mathbb R^{3}}\hat{f}_{i}(-\alpha v^{2}+vd_{i}+ l_{i} U(x,t))dv,
\end{equation}
and we assume that $U\in C^{2}(\bar{\Omega})$ is the solution of boundary problem 
\begin{equation}
\left \{\begin{array}{c}
LU\stackrel{\triangle}{=}\frac{\partial^{2}U}{\partial t^{2}} - c^{2}\triangle U= h_{f}(\cdot, U)\;\;{\rm in}\; \Omega,\\[0.2cm]
U=0, \;\;{\rm on}\;\;\;\partial\Omega.
\end{array}\right.                           
\end{equation}
We define 
$$
U(x,t)= \tilde{U}(x+ td), \;\;\; t\ge 0, \;\;\; x\in\bar{\Omega},
$$
$$
K(x,t)\stackrel{\triangle}{=}-\frac{d^{2}}{4\alpha^{2}c^{2}- d^{2}} \left (\nabla U(x,t) - |d|^{-2}\partial_{t} U(x,t) d\right ), \;\;\;t\ge 0, \;\;\;x\in \bar{\Omega},
$$
$K\in C^{1}([0,\infty[\times \Omega)^{3}$ and $E, B$ by means of (23), (24).   
Then, $(f_{1},\ldots, f_{n}, E, B)$ is a classical solution of the VM system in $\Omega$, and it satisfies boundary conditions (25) and (26)}.
\end{lemma}

{\it Proof.} According to Lemma 2, $h_{f}$ is a continuous and continuously differentiable function. Function $U$ satisfies (9). Therefore, it follows from Theorem 1 that $f_{1}, \ldots, f_{n}$ is a solution 
of the Vlasov equation, and $E, B$ is a solution of the Maxwell system. As $U$ vanishes on $\partial\Omega$, from the definition of $U$ and the translation invariance $\Omega$ in $d$, we obtain that $U$ and $\partial_{t}U$ vanish on $[0,\infty[\times \partial\Omega$. Hence, $\nabla U\times n_{\Omega}=K\times n_{\Omega}=0$ on $[0,\infty[\times \partial\Omega$. From last one, we obtain 
$$
E(x,t)\times n_{\Omega}(x)= (K(x,t) - \nabla U(x,t))\times n_{\Omega}(x)=0
$$
and 
$$
B(x,t)\times n_{\Omega}(x)= |d|^{-2}(n_{\Omega}(x)\times K(x,t))d=0
$$
at $t\ge 0$ and $x\in \partial\Omega$. Therefore, the boundary conditions (25) are satisfied.               \hspace{1.5cm}         $\Box$

\begin{theorem} 
{\it Let $\Omega \subset {\mathbb R}^{3}$. Let $f_{1}, \ldots, f_{n}:\; {\mathbb R}\rightarrow [0,\infty[$ satisfy condition ${\bf (f1)}$ and be (pointwise) less than the corresponding functions $\psi_{1}, \ldots, \psi_{n}:\; {\mathbb R}\rightarrow [0,\infty[$ satisfying condition ${\bf (f4)}$ with $\mu>0$. We suppose that $|d|<1$ and there exists function $\tilde{U}\in C_{C}^{2}(\Omega)$ such that 
$$
U(x,t)= \tilde{U}(x+ td), \;\;\; t\ge 0, \;\;\; x\in \Omega. $$

Then, (29) in Lemma 3 possesses a smooth solution, and $f_{1}, \ldots, f_{n}$ generates the classical solution:

$$(f_{1}, \ldots, f_{n}, E, B)$$ of the VM system in $\Omega$ in the form (22)--(24)}.
\end{theorem}

{\it Proof.} Since elliptic operator $L$ in (29) has constant coefficients, then through the linear change of coordinates, it is possible to transform it to Laplace operator $L= \triangle$. We introduce notations $F\stackrel{\triangle}{=}
(f_{1},\ldots, f_{n})$, and express the right-side of $h_{F}$ of (28) as 
$$
h_{F}(x,U)= -c_{1}(c_{2}- d_{2}) \sum_{i=1}^{2}q_{i}h_{f_{i}} (l_{i}U(x)),
$$
where functions $h_{f_{1}}, \ldots, h_{f_{n}}$ are defined in Lemma 3. From Lemmas 2 and 3 we obtain
$$
h_{F}(x,U)\left \{\begin{array}{c}
\ge -c_{1}(c_{2}-d_{2}) \sum_{q_{i}>0} C_{i}|q_{i}| h_{\psi_{i}}(|l_{i}| \tilde{U}(x))\stackrel{\triangle}{=}h_{1}(x,U),\\[0.2cm]
\le c_{1}(c_{2}- d_{2})\sum_{q_{i}<0}C_{i}|q_{i}|(-|l_{i}|\tilde{U}(x))\stackrel{\triangle}{=}h_{2}(x, U),   \hspace{0.5cm}
\end{array}\right.
$$
where $h_{\psi_{1}},\ldots , h_{\psi_{n}}:\; {\mathbb R}\rightarrow {\mathbb R}$ - continuously differentiable, decreasing, and non-negative functions. Moreover functions $h_{1}, h_{2}$ are continuously 
differentiable and increasing in $U$ and $h_{1}\le 0\le h_{2}$.

\section{Numerical illustration}

In this section, we demonstrate the propagation of a non-linear travelling wavefront of the self-consistent electromagnetic field potential $U(x,t)$ within domain $\Omega$. Mathematically, the transport equation describes the non-stationary transport of charged particles in the domain $\Omega$, and the boundary conditions move in time similar to a non-linear wave. We verify this through computer simulation within a complex 3D model, and compute the numerical solutions of the 3D hyperbolic sh-Gordon equation for the self-consistent electromagnetic field potential $U(x,t)$ derived in the previous sections. 

For this numerical illustration, we solve 
\begin{equation}
\Delta u = \sinh(u) \quad \mbox{ in } \Omega
\end{equation}
with the Dirichlet boundary conditions given by 
$u(x)=x_1$ and also $u(x)=\sin(x_1+x_2+x_3)$. Domain $\Omega$ has been defined in the previous sections. We run the simulation through the truncated cone z=0 to Z=L. 

For the numerical computations, we utilize the finite element library 
FEniCS. For the domain geometry and mesh, we use FreeCAD and Gmsh software. See \cite{AlnaesBlechta2015a,LoggMardalEtAl2012a, geuzaine2009gmsh, gayer2016freecad}. Visualization is realized with ParaView, see \cite{squillacote2007paraview}

\begin{figure}
\center{
\includegraphics[height=0.4\textwidth]{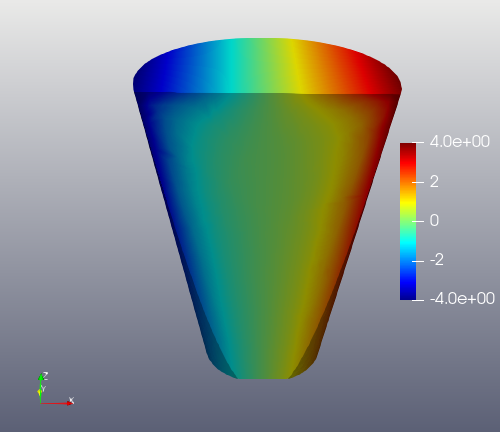}\quad
\includegraphics[height=0.4\textwidth]{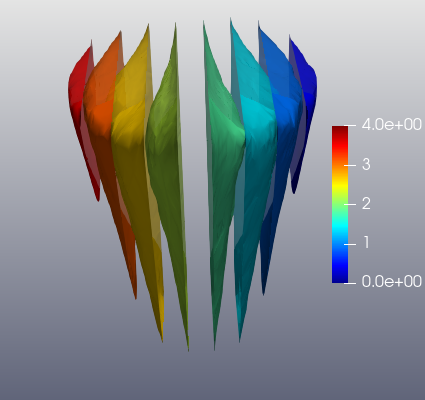}}
\caption{Solution with Dirichlet boundary condition $u(x)=x_1$ (left). Some of the level sets (right)}
\label{figdes}
\end{figure}

\begin{figure}
\center{
\includegraphics[height=0.4\textwidth]{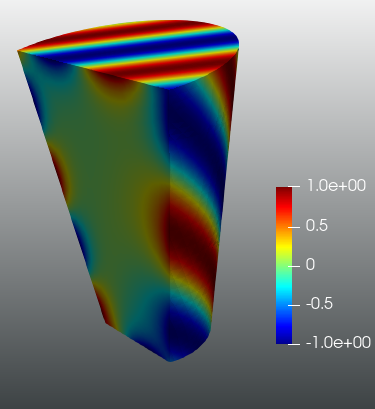}\quad
\includegraphics[height=0.4\textwidth]{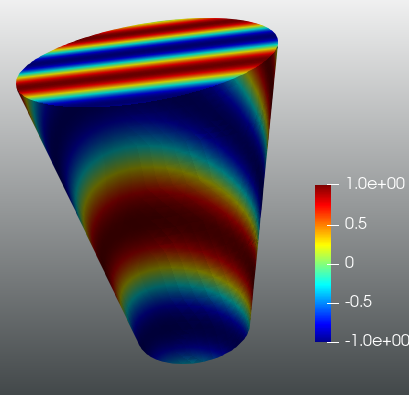}}
\caption{Solution with Dirichlet boundary condition $u(x)=\sin(x_1+x_2+x_3)$.}
\label{figdes2}
\end{figure}

The Newton method convergence of the FEniCS automatic differentiation is indicated in Table 1, and the solution is visualization in Figures \ref{figdes} and \ref{figdes2}. 

\begin{table}[]
\begin{verbatim}
    Solving nonlinear variational problem.
  Newton iteration 0: r (abs) = 2.835e+03 (tol = 1.000e-10) r (rel) = 1.000e+00 
  Newton iteration 1: r (abs) = 2.050e+00 (tol = 1.000e-10) r (rel) = 7.231e-04 
  Newton iteration 2: r (abs) = 9.833e-02 (tol = 1.000e-10) r (rel) = 3.468e-05 
  Newton iteration 3: r (abs) = 4.909e-04 (tol = 1.000e-10) r (rel) = 1.732e-07
  Newton solver finished in 3 iterations and 3 linear solver iterations.

\end{verbatim}

    \caption{Output of the FEniCS solve command.}
    \label{tab:my_label}
\end{table}

\newpage

\subsection{Second-order hyperbolic sh-Gordon equation}
In this section, we present a numerical example for the solution of 
\begin{equation}
\frac{d^2 u}{dt^2}-\Delta u = \sinh(u) \quad \mbox{ in } \Omega
\end{equation}

where $\Omega$ is the previously introduced truncated cone. 
We divide the boundary of the cone in 
$\partial \Omega=\partial_T \cup \partial_H\cup \partial_B $ for the top, hull (body of the cone), and bottom of the cone. For the boundary data, we impose 
\begin{equation}
u(t,x)={u_\partial}(t,x) \mbox{ for } x\in {\partial_B} \cup \partial H ;  \frac{du}{dt}(t,x)=\frac{du_\partial}{dt}(t,x) \mbox{ for } x\in {\partial_B}\cup \partial_H.
\end{equation}

Additionally, we impose homogeneous Neumann data on $\partial_T$. We also impose the initial potential
\begin{equation}
 u(0,x)=u_0(x) \mbox{ for } x\in D
\end{equation}
and the initial velocity 
\begin{equation}
 u(0,x)=v_0(x) \mbox{ for } x\in D.
\end{equation}

We only impose special boundary data 
$u_\partial$ that satisfies the transport equation along the boundary: 
\begin{equation}
u_\partial(t,x)= u_B(z(x,t)), 
\end{equation}
where $u_B$ is defined at $\partial_B$ and for $x\in \partial_H$; $z(x,t)\in \partial_B$ represents the intersection of the bottom-circle of the cone with a line in the cone that is also contained in the tangent plane of the cone hull at $x$. 

We follow \cite[step-23]{bangerth2007deal} for the solution of the wave equation. We introduce $v=\frac{du}{dt}$ and represent the first order (in time) system as 
\begin{equation}
\frac{du}{dt}-v=0 \mbox{ in } D; \frac{dv}{dt}-\Delta u=\sinh(u) \mbox{ in  } D,
\end{equation}
with the same boundary conditions described above. 
We then discretize time interval $[0,T]$ into $T/\delta t$ time intervals 
$t_0=0,t_1=\delta t, \dots$. Potential $u$ is denoted as $u_n:=u(t_n,\cdot)$, and velocity $v$ is denoted likewise. We then express the 0.5-Crank-Nicolson scheme in time for the linear hyperbolic part and set the non-linear term in explicit form. We obtain, 

\begin{equation}
  \begin{split}
\frac{1}{\delta t} u_{n} -0.5 v_n = \frac{1}{\delta t} u_{n-1}+0.5 v_{n-1}\\ 
\frac{1}{\delta t} v_{n} -0.5\Delta u_n=
\frac{1}{\delta t} v_{n-1} +0.5 \Delta u_{n-1}+\sinh(u_{n-1})
\end{split}.
\end{equation}
 
Substituting $v_n$ from the second equation into the first, we obtain the following system:
\begin{equation}
  \begin{split}
u_{n} -0.25\delta t^2 \Delta u_n =0.5 v_{n-1}+u^{n-1}+0.25\delta t^2\Delta u_{n-1}+0.5\delta t^2 \sinh(u_{n-1}) \\
v_{n} -0.5\delta t\Delta u_n=v_{n-1} +0.5\delta t \Delta u_{n-1}+\delta t\sinh(u_{n-1}).
\end{split}
\end{equation}

We then perform Galerkin projection of these equations into finite element space considering the previously explained boundary conditions. 

\subsection{Numerical illustration (Case I)}

\begin{figure}
\center{
\includegraphics[height=0.2\textwidth]{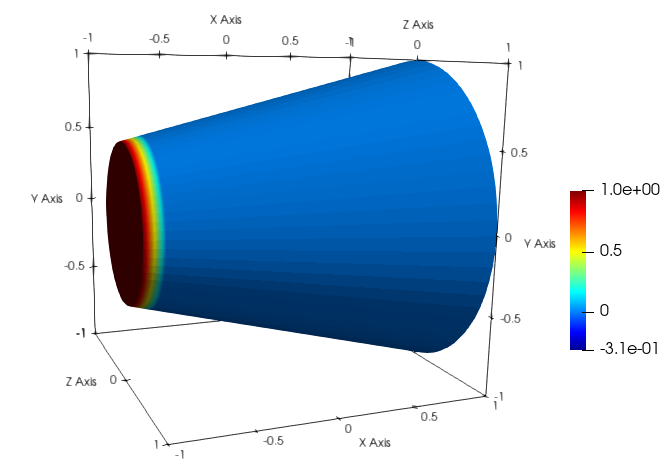}\quad
\includegraphics[height=0.2\textwidth]{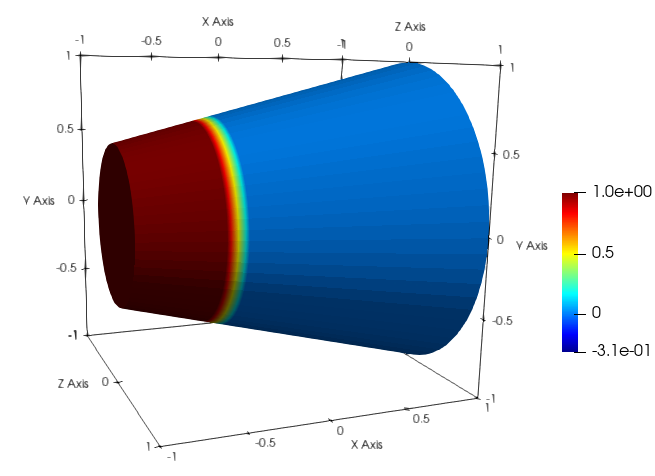}
\includegraphics[height=0.2\textwidth]{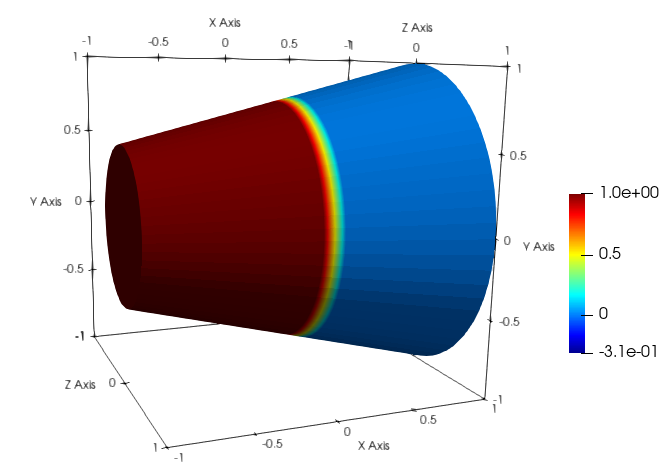}
}
\center{a)\phantom{xxxxxxxxxxxxxxxxxxxxx}b)\phantom{xxxxxxxxxxxxxxxxxxxxx}c)}
\caption{Solution of the transport equation in the boundary at $t=0.5,1,1.5,2$}
\label{figTRANSPORT}
\end{figure}

\begin{figure}
\center{
\includegraphics[height=0.2\textwidth]{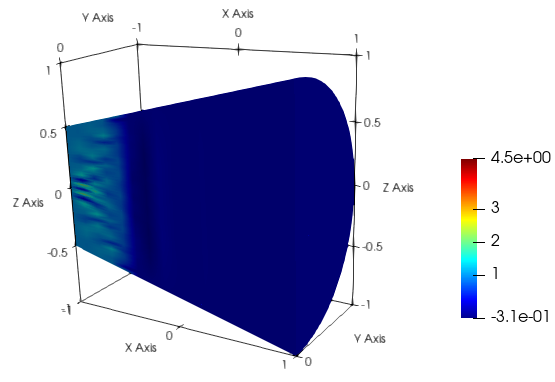}\quad
\includegraphics[height=0.2\textwidth]{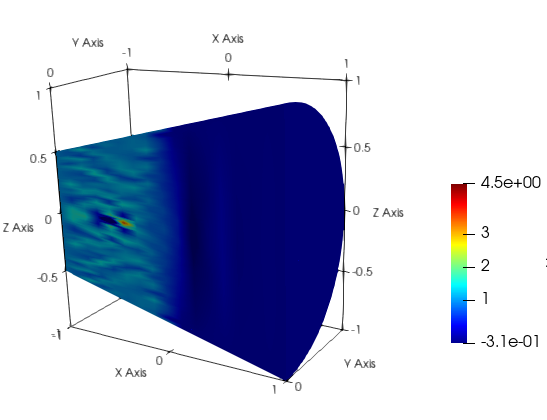}
\includegraphics[height=0.2\textwidth]{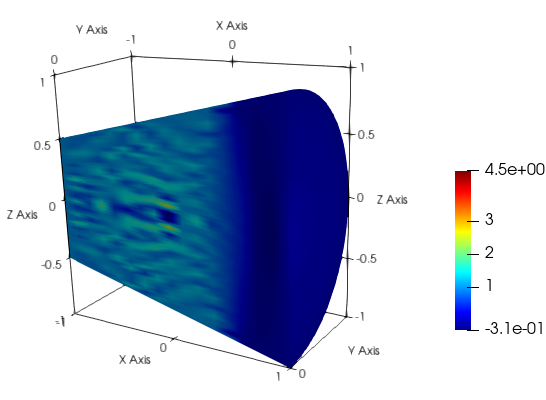}
}
\center{a)\phantom{xxxxxxxxxxxxxxxxxxxxx}b)\phantom{xxxxxxxxxxxxxxxxxxxxx}c)}
\center{
\includegraphics[height=0.2\textwidth]{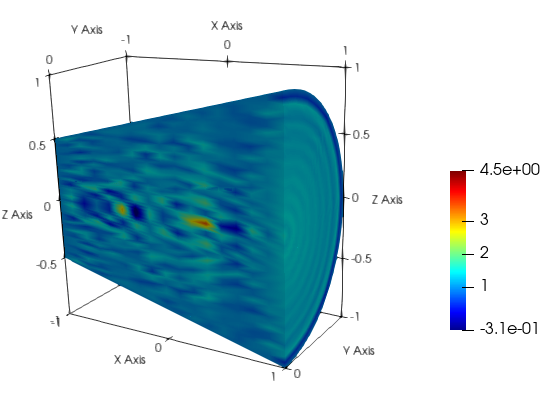}\quad
\includegraphics[height=0.2\textwidth]{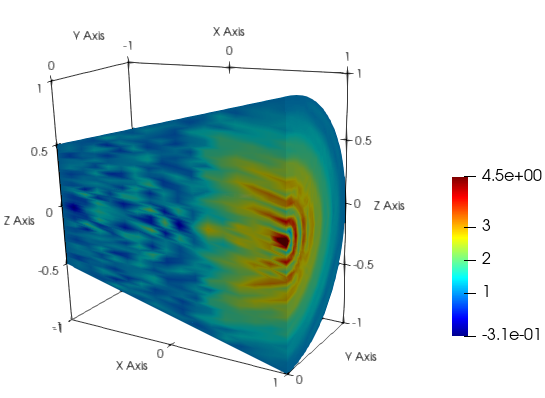}
\includegraphics[height=0.2\textwidth]{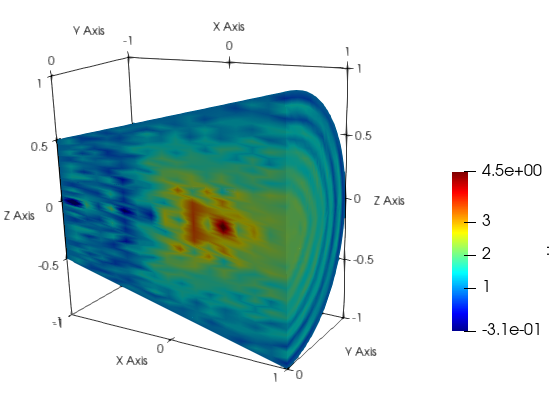}
}
\center{d)\phantom{xxxxxxxxxxxxxxxxxxxxx}e)\phantom{xxxxxxxxxxxxxxxxxxxxx}f)}

\caption{Solution of the wave equation at $t=1,1.5,2,2.5,3,3.5$}
\label{figWAVE}
\end{figure}

Throughout the numerical simulation, we consider potential $u_\partial=1$ at the contact $\Omega_{0}$ of domain $\Omega$. This value is transported along the boundary with a velocity that is positive vertically upwards (along the $y-$axis), as shown in Figure \ref{figTRANSPORT}. The solution of the wave equation is depicted in Figure \ref{figWAVE}. As a result, the travelling wave in domain $\Omega$ becomes decoupled from the potential at $\Omega_{0}$, allowing the dynamics in the domain $\Omega$ to propagate. This approach corresponds to the commonly used experimental conditions.

When the travelling wave moves in time and approaches the top ($\partial_T$) of the truncated cone as shown in Figures\ref{figWAVE}a, \ref{figWAVE}b, and \ref{figWAVE}c, the charged particle density along the vertical axis increases, generating a precursory filamentary structure of the enhanced self-consistent electromagnetic field potential $U(x,t)$. Its typical form is shown in Figure \ref{figWAVE}d. The value of the potential is apparently maximum on the vertical axis in the upper-middle part of domain $\Omega$.  

When the travelling wave reaches the top of the truncated cone at $\partial_T$ of domain $\Omega$, the value of the potential strongly increases in the wavefront. Further increase of the potential results in complete reconfiguration of the spatial modulated filaments as observed in Figure \ref{figWAVE}e. Non-stationary transport of the charged particles continues to occur synchronously at the top ($\partial_T$) of domain $\Omega$. The charged particle density increases, and a stable stationary structure with distinct boundaries is formed. At the top ($\partial_T$) of domain $\Omega$, nested coaxial-like multiple-walled tubes emerge strictly along the vertical symmetry axis. Each ring-like tube is characterized by the value of the self-consistent electromagnetic field potential. The potential decreases in the radial direction, with the least value farthest from the symmetry axis;   potential $U(x,t)$ decays in the direction of the electric field. In this regime, no filamentary structures exist in the lower-part of domain $\Omega$. Its typical form is shown in Figure \ref{figWAVE}e. 

In a longer time-interval, the periodic nested hierarchies transform into an apparent chaotic spatial scenario. Its typical form is shown in Figure \ref{figWAVE}f. This regime is characterized by an apparent maximum of the self-consistent electromagnetic field potential $U(x,t)$ along the vertical axis of symmetry in the middle of domain $\Omega$. In this regime, no periodicity exists in domain $\Omega$.

\subsection{Numerical illustration (Case II)}

Throughout the numerical simulation, domain $\Omega$ is controlled by the potential at contact space $\Omega_{0}$:  

\begin{equation}
u_\partial(t)=\left\{\begin{array}{cc}
    1 &  0\leq t\leq 0.2 \\
    0 &  \mbox{otherwise},
\end{array}\right.
\end{equation}

\begin{figure}
\center{
\includegraphics[height=0.2\textwidth]{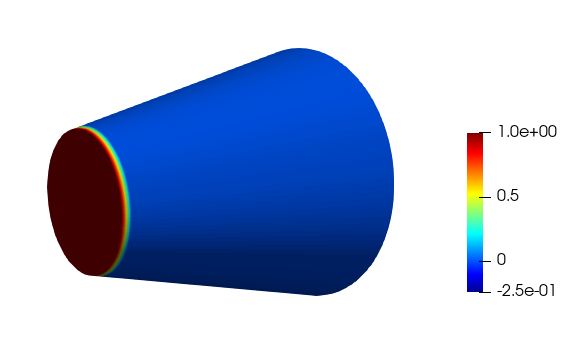}\quad
\includegraphics[height=0.2\textwidth]{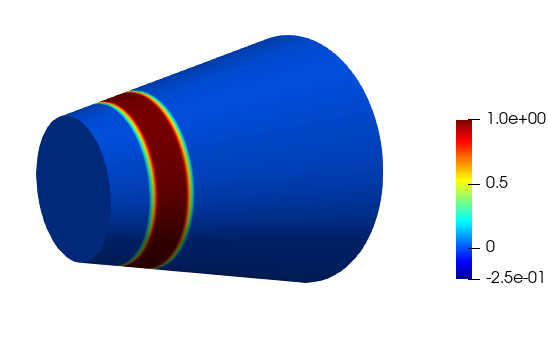}
\includegraphics[height=0.2\textwidth]{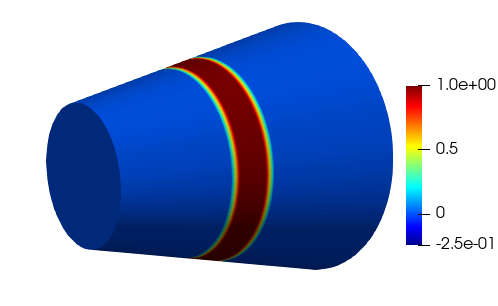}
}
\center{a)\phantom{xxxxxxxxxxxxxxxxxxxxx}b)\phantom{xxxxxxxxxxxxxxxxxxxxx}c)}
\caption{Solution of the transport equation in the boundary at $t=0,0.5,1,1.5,2$.}
\label{figTRANSPORT2}
\end{figure}

This value is transported along the boundary with a velocity that is positive
vertically upwards (along the $y-$axis), as shown in Figure \ref{figTRANSPORT2}. The solution of the wave equation is depicted in Figure \ref{figWAVE2}. An illustrative video is presented in the Supplementary material. 

\begin{figure}
%\\center{
%\\includegraphics[height=0.2\textwidth]{step10.png}
%\\quad
%\\includegraphics[height=0.2\textwidth]{step15.png}
%\}
%\\center{a)\phantom{xxxxxxxxxxxxxxxxxxxxx}b)}
\center{
\includegraphics[height=0.35\textwidth]{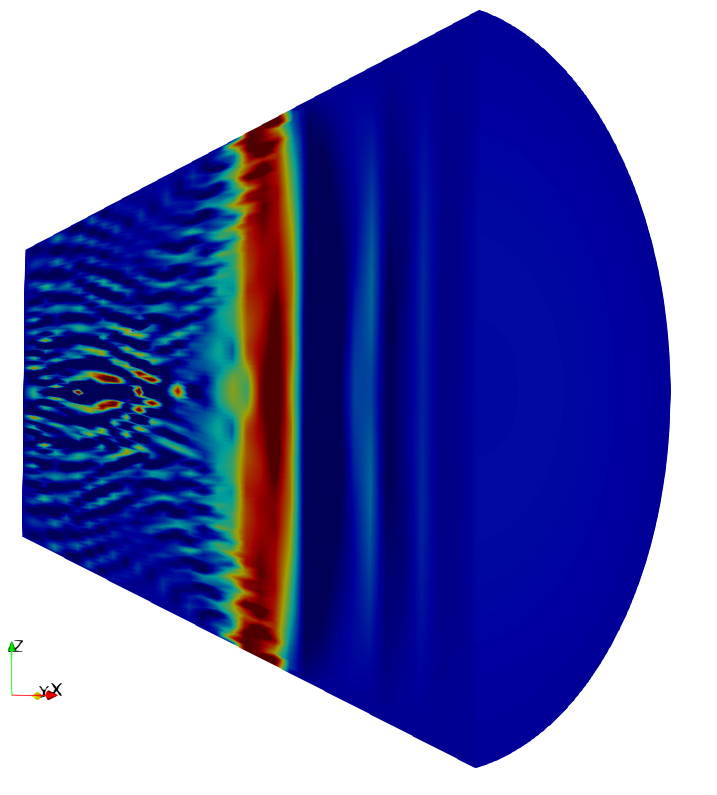}\quad
\includegraphics[height=0.35\textwidth]{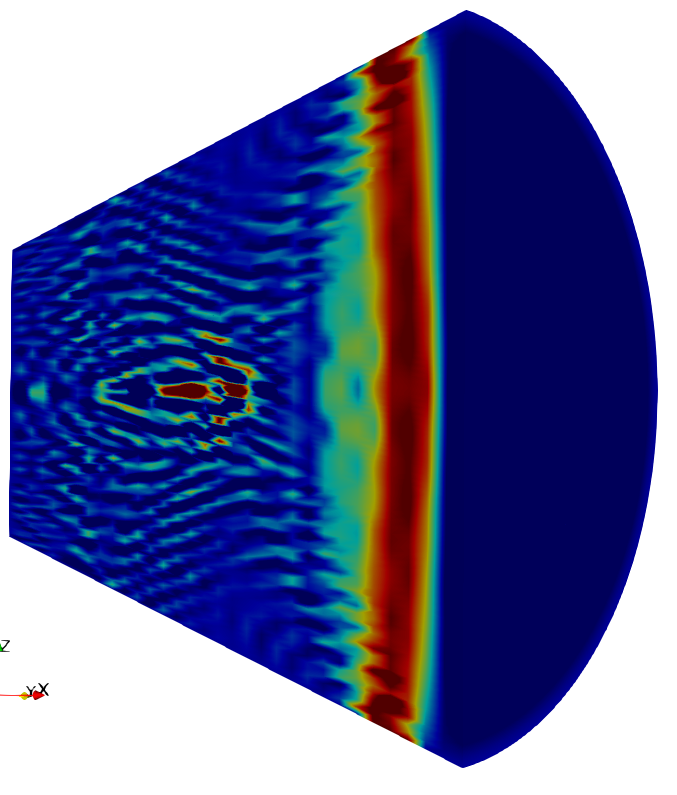}
\includegraphics[height=0.35\textwidth]{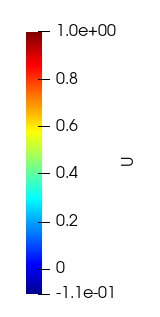}
}
\center{a)\phantom{xxxxxxxxxxxxxxxxxxxxx}b)}

\center{
\includegraphics[height=0.35\textwidth]{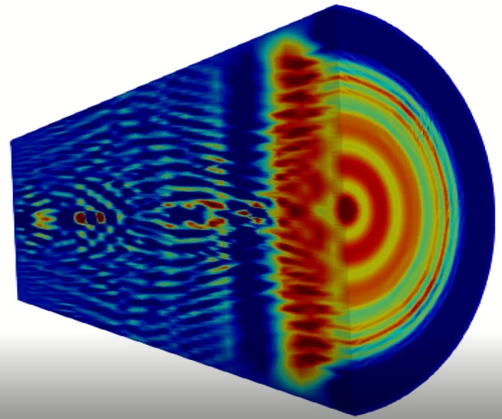}\quad
\includegraphics[height=0.35\textwidth]{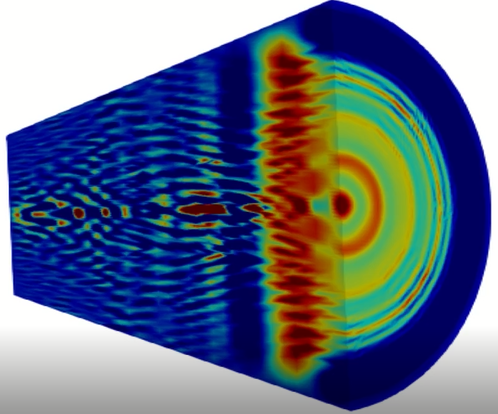}
}
\center{c)\phantom{xxxxxxxxxxxxxxxxxxxxx}d)}

\center{
\includegraphics[height=0.35\textwidth]{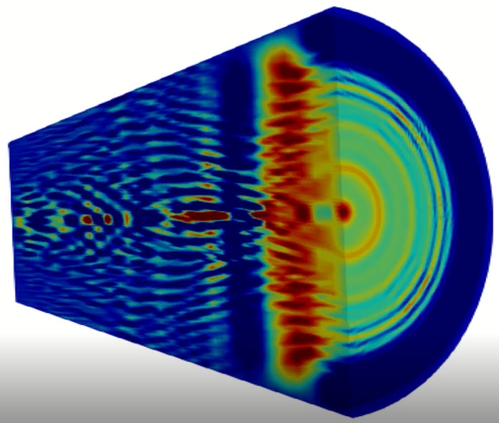}\quad
\includegraphics[height=0.35\textwidth]{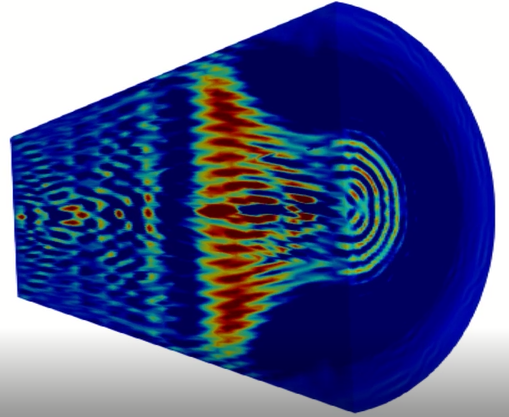}
}
\center{e)\phantom{xxxxxxxxxxxxxxxxxxxxx}f)}

\caption{Solution of the wave hyperbolic sh-Gordon equation at $t=0.2,0.25,0.3,0.35,0.4,0.45$ from top-to-bottom and left-to-right. }
\label{figWAVE2}
\end{figure}

It can be shown that the charged particle density and consequently, the value of potential $U(x,t)$ are maximum at the wavefront, when the travelling wave propagates through domain $\Omega$, as depicted in Figures \ref{figWAVE2}a and \ref{figWAVE2}b. In contrast, when the travelling wavefront reaches the upper-part of domain $\Omega$, at the top ($\partial_T$), the enhanced self-consistent electromagnetic field potential $U(x,t)$ becomes non-linear. As shown in Figures \ref{figWAVE2}c, d, and e, the non-linear travelling wavefront transforms into a spot on the vertical axis. The circular rings evolve concurrently with radial periodicity to the central spot. Each ring is characterized by the value $U(x,t)$ of the self-consistent electromagnetic field. During this transformation, the charged particle density along the vertical axis retains the precursory filamentary structure of the enhanced self-consistent electromagnetic field potential $U(x,t)$ in the rest of domain $\Omega$. It is to be noted that no regular modulation of the longitudinal and radial spatial structures exists in the small space between the top and upper-middle part of domain $\Omega$. Its typical form is shown in Figures \ref{figWAVE2}c, d, and e. As depicted in Figure \ref{figWAVE2}f, the propagation of the non-linear travelling wave (case II) results in nested hierarchies, and is bent in the direction of the Lorentz force. The final spatial structure becomes fixed in the domain, and forms two completely distinctive configurations. The coupling between the two regions is most intriguing.

It can be shown that the formation of the two configurations and the most- likely formation of nested hierarchies satisfy the phase-matching resonance, i.e. , coupling occurs through the phase and amplitude of the oscillating waves. Through discrete Fourier transform (DFT) of the numerical data of the amplitude of potential $U(x,t)$ , the values of the frequencies can be determined. 

\begin{figure}
\center{
\includegraphics[height=0.2\textwidth]{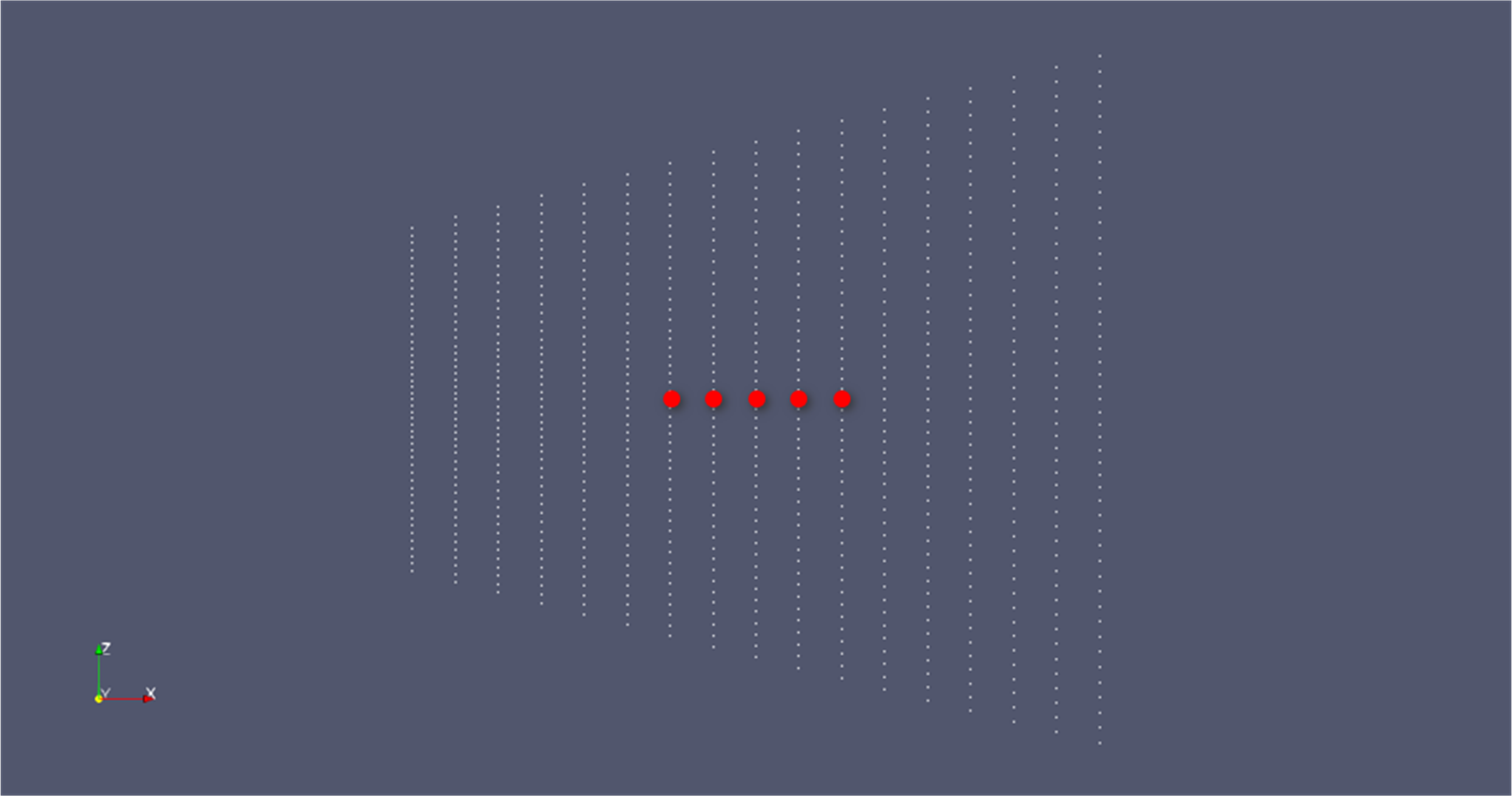}\quad
\includegraphics[height=0.2\textwidth]{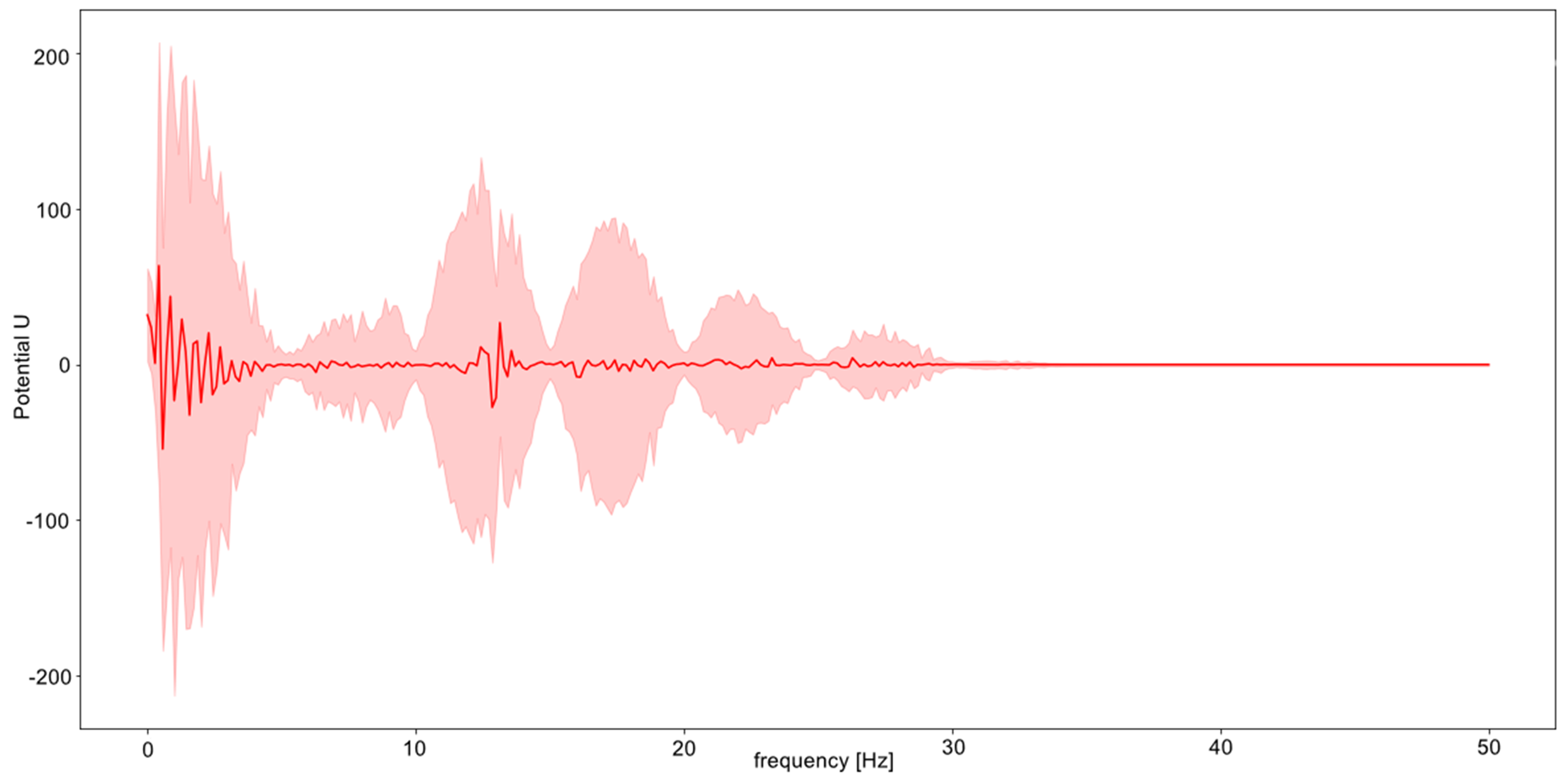}
}
\center{a)\phantom{xxxxxxxxxxxxxxxx}b)}
\center{
\includegraphics[height=0.2\textwidth]{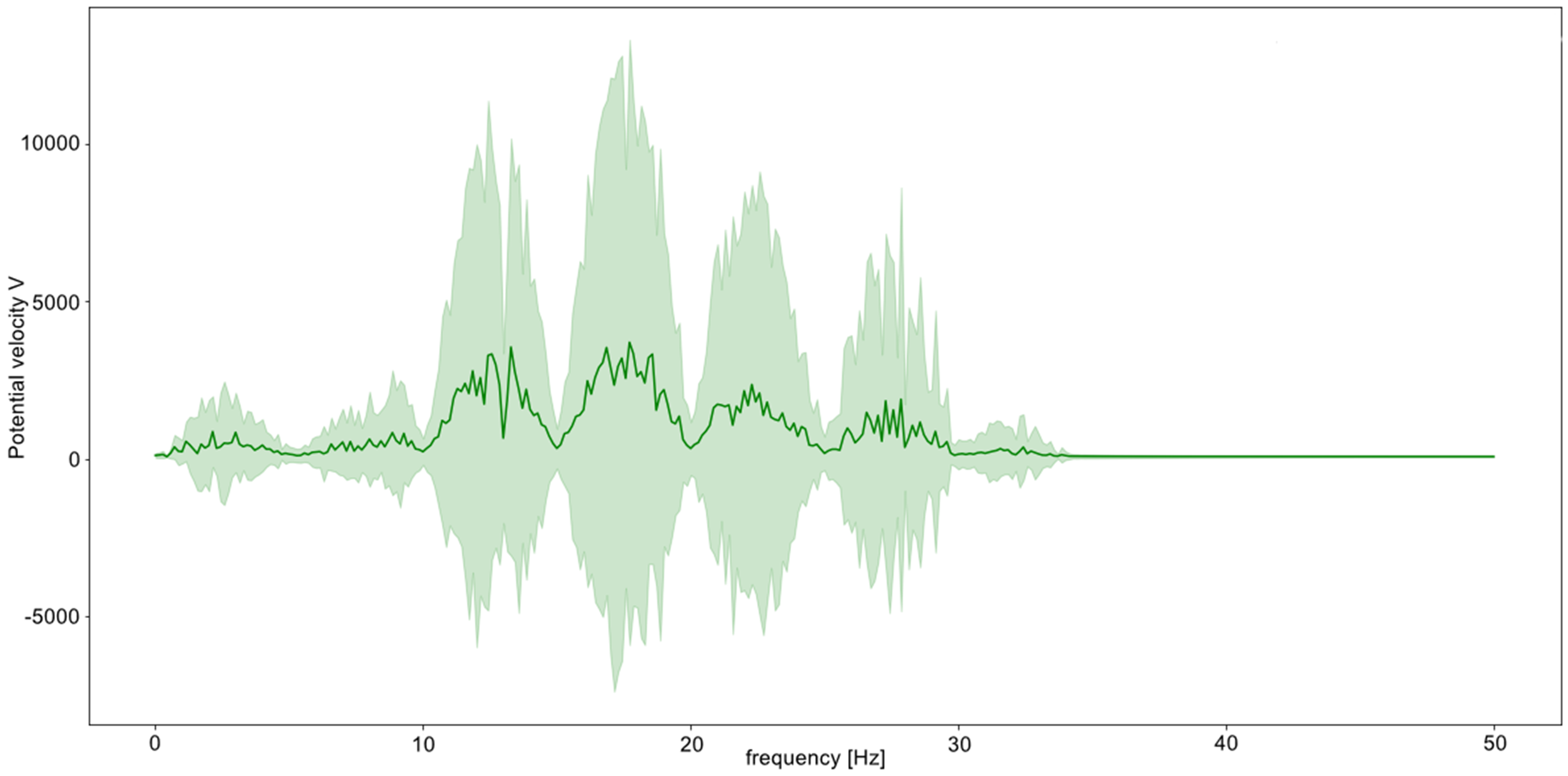}\quad
\includegraphics[height=0.2\textwidth]{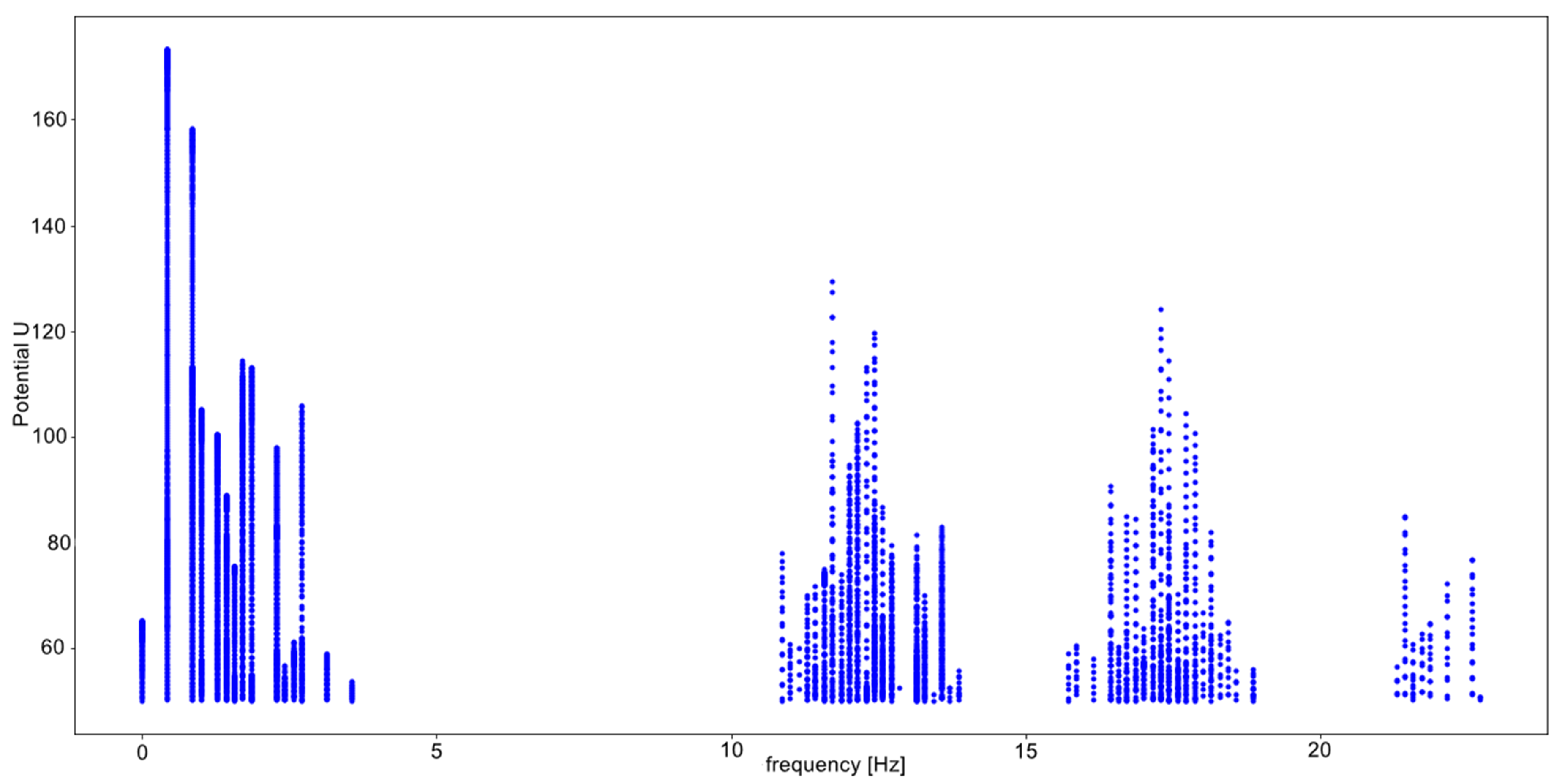}
}
\center{c)\phantom{xxxxxxxxxxxxxxxx}d)}
\caption{a) Selected points along the revolution axis of the cone. b) Mean ensemble-based spectral signal and reported confidence margin representation of the potential of the frequency component. c) Mean ensemble-based spectral signal and reported confidence margin representation of the frequency of the velocity component. d) Spectrum $u(f)$ of amplitude $u(t)$ with the application of a high-frequency filter with a threshold of 50.}
\label{figfreq}
\end{figure}
\newpage

To assess the efficiency of the wave mixing process, a set of finite signals are extracted from selected points along the vertical axis of the cone's symmetry as shown in Figure \ref{figfreq}a. As the time series of potential $U(x,t)$ is in real space, the DFT \cite{cooley1965algorithm} results have conjugate symmetry; thus, the imaginary part of the data is the complex conjugate of the real frequency data. Therefore, the DFT between \(0\) and \(N/2\) represents the positive frequency data for \(N\) time-steps. 

Figures \ref{figfreq}b and \ref{figfreq}c display the numerical data of the potential of the frequency component and the frequency of the velocity component, respectively. Analytically, the signal domain is first transformed from time-based to frequency-based. The numerical solver generates the time series of potential U(x, t); the data set includes discrete samples in both time and space. DFT analysis of the discrete signal is performed over k set of points:

\begin{equation}
    u(f) = \sum_{n=0}^{N-1} u_n\cdot e^{ \frac{-2\pi i}{N}k_n }.
\end{equation}

Thereby, the peaks of the spectrum evaluated with a confidence interval of \(\pm 2\cdot \sigma(u_k(f))\) are located(shaded region in Figures \ref{figfreq}b and c). To increase the accuracy, Fourier transform is performed in the vicinity of the peaks (continuous lines) with the application of a high-frequency filter. A threshold of 50 (U) is applied for the resulting set of potentials of the frequency components. 

\begin{figure}
\center{
\includegraphics[height=0.14\textwidth]{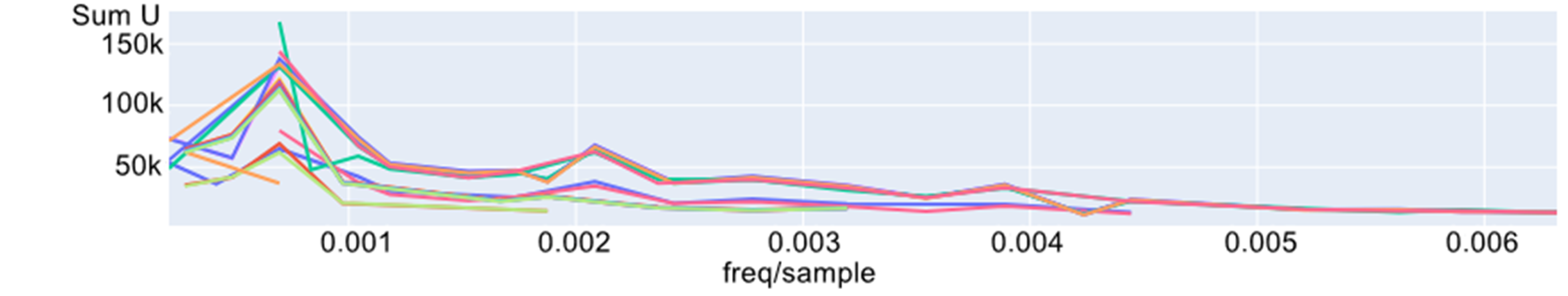}
}
\center{a)10-day interval from 10-th April 2013}
\center{
\includegraphics[height=0.12\textwidth]{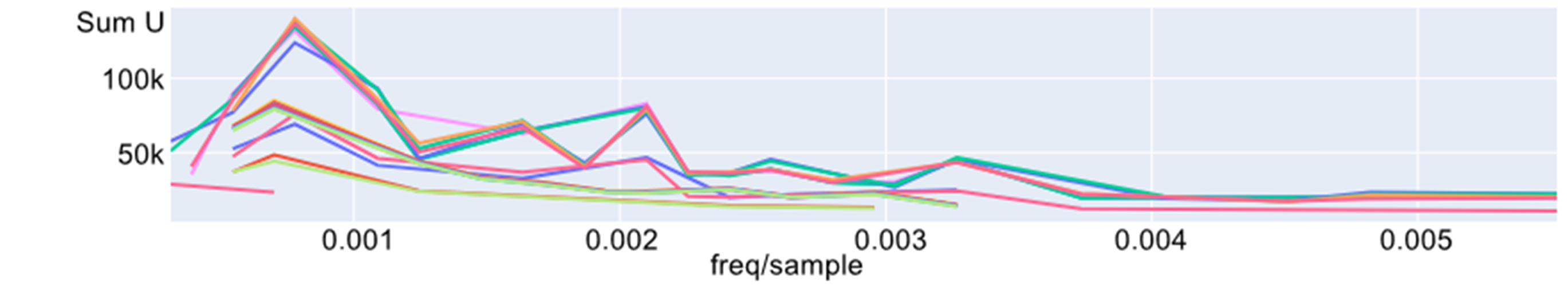}
}
\center{b)10-day interval from 20-th April 2013}
\center{
\includegraphics[height=0.12\textwidth]{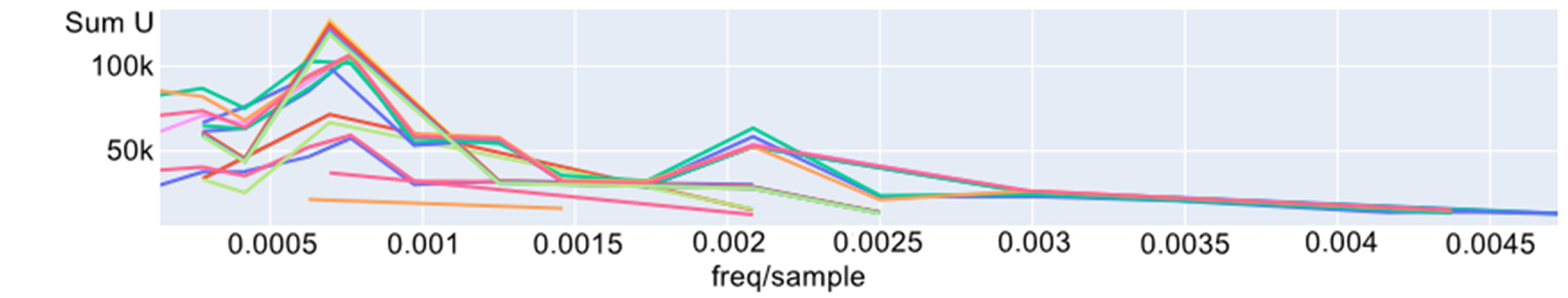}
}
\center{c)10-day interval from 30-th April 2013}
\center{
\includegraphics[height=0.12\textwidth]{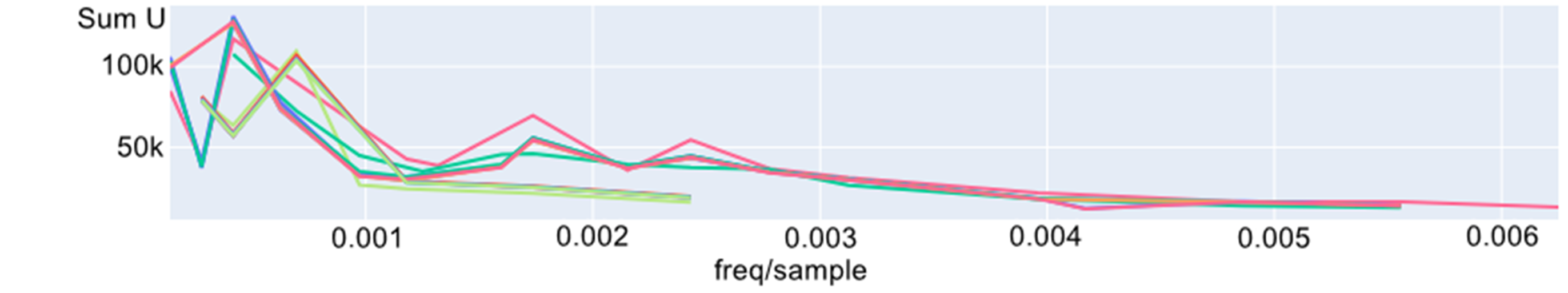}
}
\center{d)10-day interval from 10-th May 2013}
\center{
\includegraphics[height=0.12\textwidth]{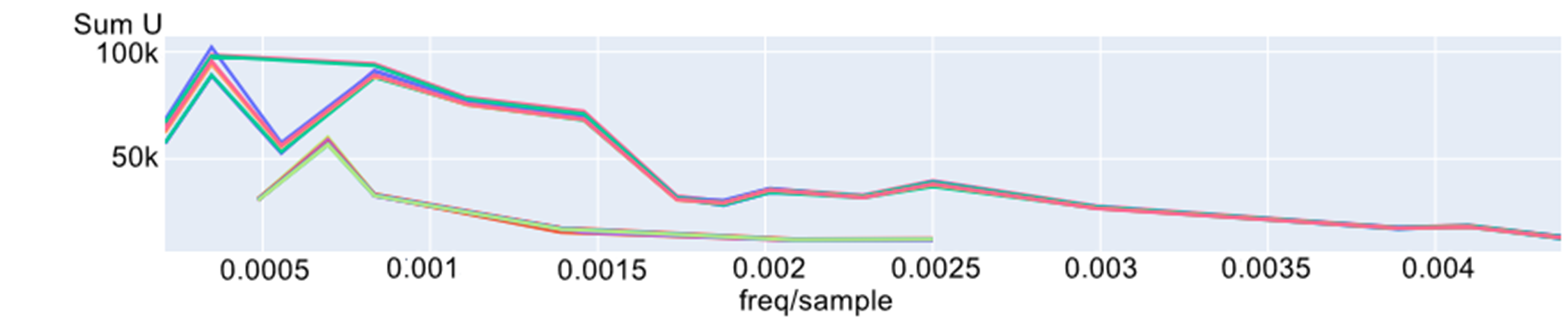}
}
\center{e)10-day interval from 20-th May 2013}
\center{
\includegraphics[height=0.12\textwidth]{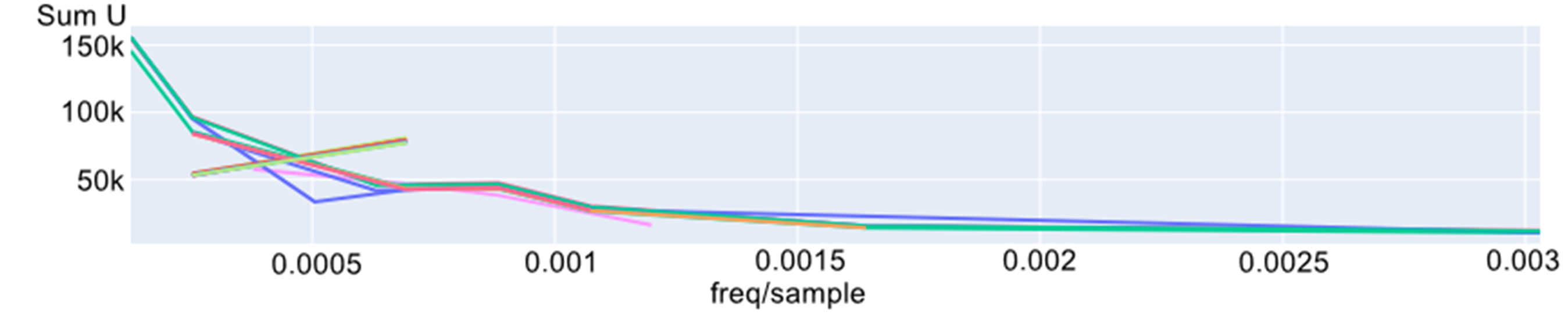}
}
\center{f)10-day interval from 30-th May 2013}
\center{
\includegraphics[height=0.12\textwidth]{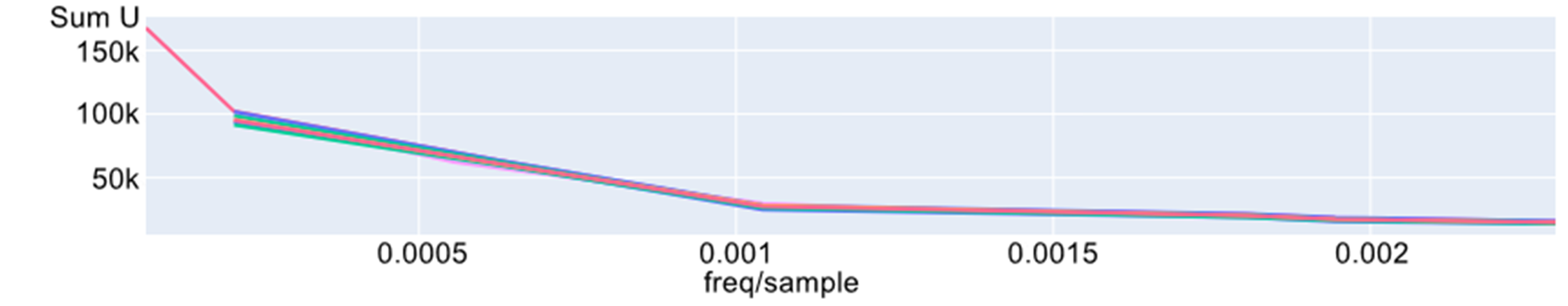}
}
\center{g)10-day interval from 10-th June 2013}
\caption{Fourier transform of the electric potential recorded at station-No4-IMFSET, located 362 km from the epi-centre of large mantle earthquake M8.3 that occurred on 24-th May 2013. The seven plots are for ten-day intervals each from 10-th April 2013 to 20-th June 2013. Sum U indicates the sum of the electric potential computed in a two-day sliding window for six channels at station-No4-IMFSET. Three frequency components appear before earthquake M8.3 and decay from the moment of its occurrence.}
\label{figfreq_M8}
\end{figure}
\newpage

The amplitudes of the components at frequencies $f$ are shown in Figure \ref{figfreq}d. The frequencies are $f_{1}$= 2 Hz, $f_{2}$= 13 Hz, $f_{3}$= 17 Hz, and $f_{4}$= 23 Hz. Certain parameter values at which the frequencies satisfy the resonance condition need to be investigated further. In this study, we compare the field measurements of the electric potential (Kuznetsov method) \cite{kuznetsov1991practice, bobrovskiy2017nonstationary} and the numerical results.  

Note that although there are the four frequency components at frequencies $f_{1}-f_{4}$, the two peaks at frequencies $f_{1}$ and $f_{2}$ represent the spectral line series. These are the main and second spectral lines, responsible for exchange between the non-linear wavefront propagating within the domain and the transport of the boundary conditions in the form of a non-linear wave that varies with time. The two peaks at frequencies $f_{3}$ and $f_{4}$ describe the wave-mixing effects, and are in resonance with any two components at frequencies $f_{1}$ and $f_{2}$. 

\begin{figure}
    \center 
    \includegraphics[height=0.35\textwidth]{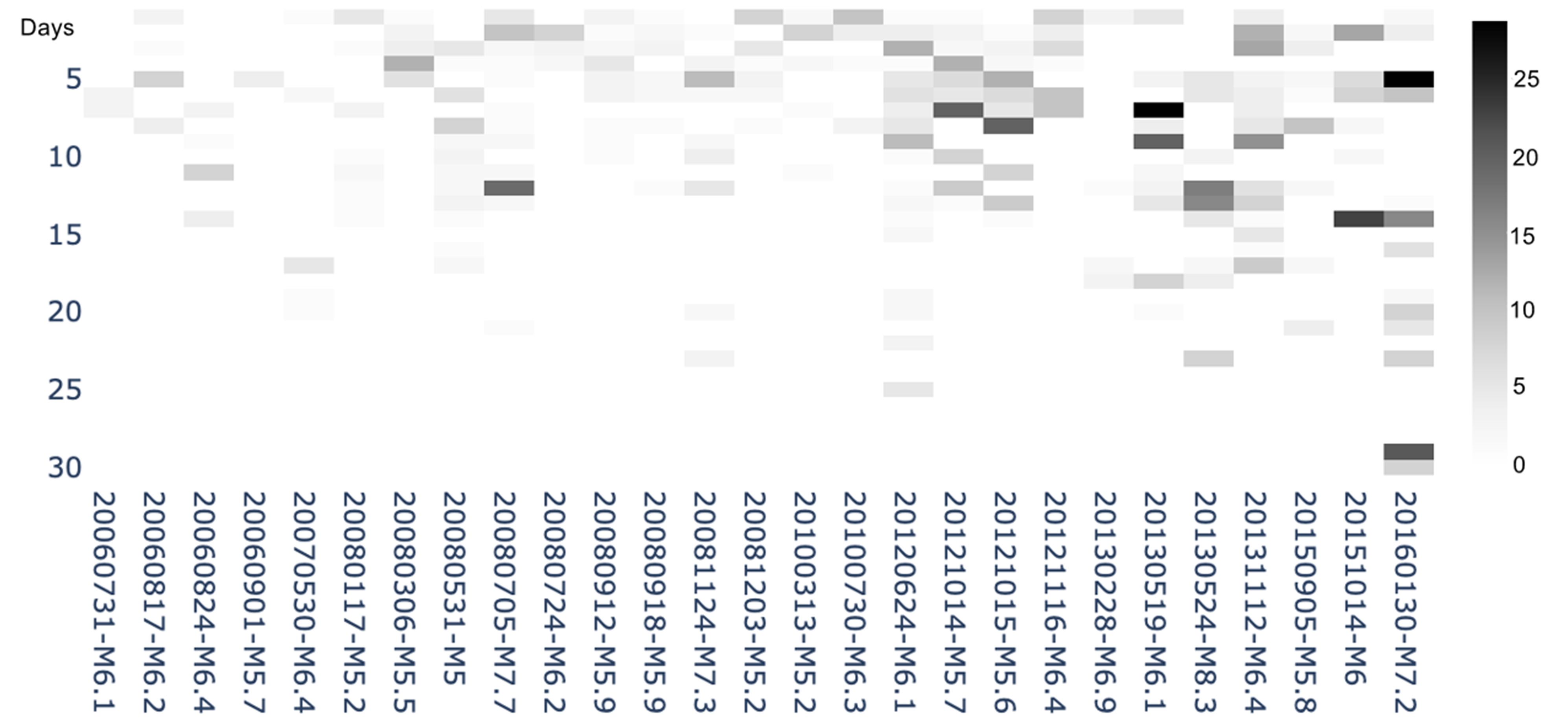}
    \caption{Aggregated sum of the electric potential computed for a group of four stations in Kamchatka and the seismic process in Kamchatka. Each station is located at 0-5-10-25 km from the reference point at 53.05° 158.65°. This reference point is located 100 km from the epi-centre of the large earthquake M7.2, which occurred on 30-th January 2016. The plots are for a 10-year interval from 01-st July 2006 to 30-th January 2016. The OY axis on the left denotes the number of days during which the changes in the electrical potential were recorded before each earthquake. The OY axis on the right is the aggregated intensity of the signals vs stations. The aggregated sum of the electric potential is maximum before earthquake M7.2 on 30-th January 2016.}
    \label{kamheatmap1}
\end{figure}

Figure \ref{figfreq_M8} compares the numerical results with the Fourier transform of the potential (V) obtained through field observations of the electric potential (Kuznetsov method). This method, which involves field observation of the electric potential, has been investigated by the extended station network (Cosmetecor) in Kamchatka, Italy, Altai, Fiji, and Sakhalin \cite{Bobrovskiy2016book, Bobrovskiy2016thesis}. This can be considered as a prototype for exploring the Earth's interior to map the effects of superionic proton conduction \cite{Hou-Nature2021}. 

We consider the large mantle earthquake M8.3, which occurred on 24-th May 2013 as an illustrative example because its occurrence depth is 598 km and Cosmetecor station No4-IMFSET is located in Esso (55.927N	158.703E), which is 362 km from the epi-centre of the earthquake. Technical details on the measurement stations are available in \cite{bobrovskiy2017nonstationary}. Quantity $U(f)$ was first computed through fast-Fourier transform and quantity $U(t)$ was computed in a two-day sliding window. These computations were repeated for six channels in total for each 10-day interval between 10-th April and 20-th June 2013. In the resultant spectrum, three peaks could be identified. The three frequency components appeared before earthquake M8.3 and decayed from the moment of its occurrence. 

As seen in Figure \ref{kamheatmap1}, the computation for a group of four stations located 0-5-10-25 km from reference point 53.05° 158.65° can reveal the spatial localization of the electric potential at the Earth's surface boundary. M7.2, the large earthquake on 30-th January 2016, was the single large event that occurred 100-km from reference point 53.05° 158.65 in a 10-year interval from 01-st July 2006 to 30-th January 2016. The accuracy of the electric potential measurements (Kuznetsov method) can be improved by considering the non-linear effect. As can be seen, the field measurements of the electric potential and the numerical results are in good agreement.

\section{Concluding remarks}

We reduced the VM system of equations to non-linear sh-Gordon hyperbolic and transport equations, which is a solution for the system of differential equations describing the dynamics over time of the distribution function consisting of two types of charged particles. In the computed 3D model, we cast a non-linear wavefront propagating within a domain and transport of the boundary conditions in the form of a non-linear wave varying with time. 

Depending on the initial potential $u_\partial$ at the bottom of domain $\Omega$, the above-mentioned non-linear wave takes the form of either nested hierarchies on a wavefront describing the process of two-wave interaction in the case where damping occurs rapidly or a wave mixing process with resonant frequencies if the phase-matching resonance is satisfied, i.e. , coupling occurs through the phase and amplitude of the oscillating waves. Certain parameter values at which the frequencies satisfy the resonance condition in case II need to be investigated further. In case II, spectrum $u(f)$ of amplitude $u(t)$ describes the spectral line series, where frequency components $f_{3}$ and $f_{4}$ are in resonance with any two components at frequencies $f_{1}$ and $f_{2}$. 

A mathematical model of proton migration in the mantle was developed, and it was concluded that the electric potential measurements (Kuznetsov method) were in good agreement with the numerical simulation results of the 3D-model. Using the sh-Gordon equation derived from the 3D wave equation, numerical simulation of a two-particle system correctly predicted the temporal features (anomalies) from the time series data (electric potential) through FFT. Large mantle (depth=598 km) earthquake M8.3 on 24-th May, 2013 was considered as an example. The numerical results showed that number of frequency components closely matched those of the field measurements. 

In case II, the final spatial structure was fixed in the domain, with two completely distinctive configurations; the nested coaxial-like multiple-walled tubes at the top of the domain showed bending in the direction of the Lorentz force. Such non-linear behaviour of the self-consistent electromagnetic field potential was in the agreement with the spatial localization of the aggregated sum of the electric potential obtained at a reference point on the Earth's surface boundary. Large earthquake M7.2 on 30-th January 2016 was the single large event that occurred 100 km from reference point 53.05° 158.65 during a 10-year interval from 01 July 2006 to 30 January 2016. The Kuznetsov method of electric potential observation and the existing station network (Cosmetecor) in Kamchatka, Italy, Altai, Fiji, and Sakhalin \cite{Bobrovskiy2016book, Bobrovskiy2016thesis} can be considered as a prototype for the exploration of free hydrogen movement in the Earth's interior. 

It is worthwhile to mention that free hydrogen (atom) migration and its collective motion can be viewed as a source that can drastically enhance the oscillator strength ($f$), which can be regarded as the effective number of electrons active in the transition \cite{Kauzmann}. The classical theory of plasma oscillations (referred to as Langmuir waves or electrostatic oscillations) dictates that harmonic oscillators (the electrons in hydrogen atoms) are permitted not only to emit electromagnetic waves but also to absorb them. If the Coulomb force that quantifies the electrostatic interaction between the traverse wave and oscillator is  Fourier transformed in a harmonic series, the resulting equation is  a differential equation of motion of a one-dimensional ideal atomic oscillator:

\begin{equation}
(\frac{d^2 x}{dt^2}) + \omega_0^{2}x = (\frac{F_\omega}{m})\cos{(\omega t+\delta_\omega)}, 
\end{equation}

where $\omega_0$ - is the cyclic frequency of free oscillation, $F_\omega$ is the frequency-dependent Coulomb force for electrostatic interaction between the traverse wave and oscillator, $\omega$ is the frequency of the traverse wave, and $\delta_\omega$ is the initial phase.

Absorption of the traverse wave will cause the ideal atomic oscillator to oscillate only if the circular frequency of the wave matches the circular frequency $\omega_0$ of the oscillator. According to classical mechanics, the frequency of oscillation of the Coulomb force [F] must coincide with the frequency of oscillation of the oscillator strength, where the latter is the force required to maintain the equilibrium position of the harmonic oscillator. When the circular frequencies of these forces are slightly diverse, broadening of the spectral absorption lines occurs. 

In quantum mechanics, the oscillator strength is a dimensionless quantity that links the theories of emission and absorption to the experiment. Formula \ref{kwant} of the oscillator strength for absorption is applied in the form:

\begin{equation}
f_{mn}= (4\pi m w/3h)g_{m}\sum|<n|x_{i}|m>|_{\alpha v^2},
\label{kwant}
\end{equation} 

where n and m are quantum numbers, h is Planck’s constant, $g_m$ is the statistical weight of the absorption levels of the atomic oscillators, and $\alpha v$ are the sum of the quantum indices of the occupied levels of the atomic oscillators. The energy difference between the two states results in broadening of the spectral line.

To summarize, as plasma carries significant free energy to excite collective oscillation, such as the well-known electrostatic Langmuir oscillation, the plasma frequency is $w_0\sim {\sqrt{n}}$. The classical Langmuir theory  offers a general interpretation of a single spectral line for the collective oscillation of charged particles, of which there is only one type in the system under consideration. In classical terms, a single harmonic oscillator model (one-component system) can be developed.

However, the experimental results for phase transition (charge-transfer state) in dense hydrogen \cite{Hanfland1993, Hemley1994}, and the recent experimental results for superionic proton conduction \cite{Hou-Nature2021, He-Nature2022} combined with the numerical results presented in this study suggest further investigation of a model of a collective of coupled harmonic oscillators (two-component system). 

The existence of non-stationary solutions for VM system of equations has been proved \cite{rudykh1989nonstationary}. Formula (\ref{nsfr}) in \cite{rudykh1989nonstationary} describes the behaviour of two types of charged particles in accordance with Vlasov's theory \cite{vlasov1961many}.

\begin{equation}
f_{i}= f_{i}(-\alpha_{i}v^2+d_{i}v+F_{i}(r,t)),
\label{nsfr}
\end{equation}

where $v$ is the velocity vector and $r$ is the radius vector of the charges, F is the force acting on a point charge, and $\alpha_{i}$ $>$0 and $d_{i}$ $\in R3$ are free parameters.

We express the first VM equation \cite{rudykh1989nonstationary} in the form (\ref{VM-osc}) of an equation representing forces. The form is close to the definition of force in Newton's second law of mechanics. 

\begin{equation}
(\frac{\partial f_{i}} {\partial t}) +v(\frac{\partial f_{i}} {\partial r})=-(\frac {q_{i}} {m_{i}})\{F_{C}+F_{L}\}(\frac{\partial f_{i}} {\partial v}),
\label{VM-osc}
\end{equation}
where m is the mass of a unit charge, and $F_{C}$ and $F_{L}$ are the Coulomb and Lorentz forces applied to the unit charge.

The first VM equation in the form (\ref{VM-osc}) enables the determination of non-stationary solutions for describing the dynamics of the  $q_1$ ($H^{+}$) condensate and $q_2$ ($OH^{-}$) condensate. More specifically, the first VM equation in the form \ref{VM-osc} enables the determination of non-stationary solutions for the forced oscillations of two-component system. 

The non-stationary solutions for the forced oscillation of two-component system, and therefore, the oscillatory strengths of two types of charged particles can be usefully addressed by the proposed PDE model. We intend to explore the experimental results \cite{Hanfland1993, Hemley1994, Hou-Nature2021, He-Nature2022} in future studies.

Appropriate treatment of the oscillatory strengths of two-component system is fundamental to correctly predict the critical pressure for superionic phase transition or the frequencies of hydrogen-dominated phonon modes in multi-component environment. The increased value of the potential can be considered as a control parameter at which resonance can occur in two-component system. This part is a research question for further investigation.  

The second important contribution of the mathematical model is that it enables better understanding of the seismic risk probabilities in a seismic-prone region in combination with the two-layer Bayesian stochastic and Neural Network model for the analysis of electric potential measurements at the Earth’s surface boundary (Bobrovskiy, paper under review) and seismic probability risk maps. The time series includes 20 years of continuous observations. For the Kamchatka simulation case, this results in the recognition of two cases with high-probability (the occurrence/non-occurrence of a large earthquake), and the probability distribution over various time frames (from months to weeks). A shortcoming of the existing approach is that most of the data has been collected in locations limited to certain cities in Kamchatka, Altai, Italy, Fiji, and Sakhalin. Some locations in the Pacific Ring such as Indonesia are more seismically active than these locations. Therefore,  coordinated effort by the scientific community is required for the continuous collection of data related to free hydrogen transport in the Earth's interior by installing stations (Kuznetsov method) in such locations. The second shortcoming is related to the existing earthquake early warning approach. In this approach, earthquake alerts are activated according to a performance-based engineering framework, and there is an engineering-based risk model to determine whether the alerts are triggered. Therefore, the current efforts of our team focus on real-time seismic analysis to include the developed mathematical model, and the two-layer Bayesian Stochastic and neural network model for various earthquake scenarios. The uncertainties that are an integral part of an earthquake warning system as well as the complex interdependence of urban lifelines and daily life need to be considered. Such interdependence necessitates appropriate seismic risk information to make decisions, against a risk metric, on the mobilization of a wide range of capacities to mitigate earthquake consequences.

\section{Methods}
\subsection*{Data availability}
The datasets generated and analysed during the current study are available in the  FigShare repository. \href{https://doi.org/10.6084/m9.figshare.19727377.v2}{Bobrovskiy, Vadim (2022): Data Time Dependent Wave. figshare. Dataset.}

\subsection*{Code availability} 
Source code of the simulation algorithm is available in FigShare repository. 

\begin{itemize}
\item \href{https://doi.org/10.6084/m9.figshare.19727308.v2}{Bobrovskiy, Vadim (2022): TimeDependentWaveonShell. figshare. Software.}

\item \href{https://doi.org/10.6084/m9.figshare.19727359.v1}{Bobrovskiy, Vadim (2022): TimeDependentWaveonCone. figshare. Software.}
\end{itemize}
Other source code is available on reasonable request to the authors.

\subsection*{Additional data} 
\begin{itemize}
\item \href{https://doi.org/10.6084/m9.figshare.19753765}{Supplementary Video 1} The nonlinear wave process in the domain $\Omega$ under potential $u_\partial=1$ at the bottom of the domain $\Omega$ (the Case 1).

\item \href{https://doi.org/10.6084/m9.figshare.19753807}{Supplementary Video 2} The nonlinear wave process in the domain $\Omega$ under potential impulse $u_\partial$ at the bottom of the domain $\Omega$ (the Case 2).

\end{itemize}

\section{Acknowledgment}

The authors are thankful to Prof. Dmitri Kuznetsov, docent and the member of International Astronomical Union and Yulianna Losyeva; Francesco Stoppa, Leonid Bogomolov, Sushil Kumar, Alex Shitov, Grigoriy Razgon, Pavel Kamenev, Alex Ganov, Alex Terentiev, Leonardo Nicoli for their support with electric potential measurements; Sergey Shopin, João Teixeira Oliveira de Menezes, Sanzhar Korganbayev, Neil Singh for the support. The authors thank the anonymous reviewers for their contribution to the peer review of this work. 

\section {Competing interests}
The authors declare no competing interests.

\section{Contributions} 
V. Bobrovskiy initiated the project, formulated the concept of the study, performed the electric potential measurements, drafted the problem statement, and wrote the manuscript with contributions from all the authors. A. Sinitsyn authored the mathematical part. J. Galbis performed the numerical simulation. M. Tognoli performed Fourier data analysis. A. Kaplin performed electric potential data analysis. P.Trucco discussed earthquake risk management.

\bibliography{references} 

\begin{thebibliography}{10}

\bibitem{Adams}
{\sc Adams, R., and Fournier, J.}
\newblock {\em Sobolev Spaces}.
\newblock Academic Press, New York, 2003.

\bibitem{AlnaesBlechta2015a}
{\sc Aln{\ae}s, M.~S., Blechta, J., and et.al.}
\newblock The fenics project version 1.5.
\newblock {\em Archive of Numerical Software 3}, 100 (2015).

\bibitem{bangerth2007deal}
{\sc Bangerth, W., Hartmann, R., and Kanschat, G.}
\newblock Deal. ii—a general-purpose object-oriented finite element library.
\newblock {\em ACM Transactions on Mathematical Software (TOMS) 33}, 4 (2007),
  24--es.

\bibitem{Bobrovskiy2016thesis}
{\sc Bobrovskiy, V.}
\newblock {\em Software and hardware of international spatially distributed
  monitoring network for investigation local and global effects prior to the
  strong earthquakes. Phd Thesis. Geophysics. (in Russian)}.
\newblock Russian State University for Geological Prospecting, Moscow, 2016.

\bibitem{Bobrovskiy2016book}
{\sc Bobrovskiy, V., and Kuznetsov, D.}
\newblock {\em Seismic global conception on the example of strongest
  earthquakes with M8+ occurred in 2001-2015 (in Russian)}.
\newblock Scientific world, Moscow, 2016.

\bibitem{bobrovskiy2017nonstationary}
{\sc Bobrovskiy, V., Stoppa, F., Nicoli, L., and Losyeva, Y.}
\newblock Nonstationary electrical activity in the tectonosphere-atmosphere
  interface retrieving by multielectrode sensors: case study of three major
  earthquakes in central italy with m 6.
\newblock {\em Earth Science Informatics 10}, 2 (2017), 269--285.

\bibitem{braasch1997semilineare}
{\sc Braasch, P.}
\newblock {\em Semilineare elliptische Differentialgleichungen und das
  Vlasov-Maxwell-System}.
\newblock Utz, Verlag Wiss., 1997.

\bibitem{cooley1965algorithm}
{\sc Cooley, J.~W., and Tukey, J.~W.}
\newblock An algorithm for the machine calculation of complex fourier series.
\newblock {\em Mathematics of computation 19}, 90 (1965), 297--301.

\bibitem{Fortuin}
{\sc Fortuin, L.}
\newblock The wave equation in a medium with a time-dependent boundary.
\newblock {\em J. Acoust. Soc. Am. 53\/} (1973), 1683--1685.

\bibitem{Gao2022}
{\sc Gao, H., Liu, C., Shi, J., Pan, S., Huang, T., Lu, X., Wang, H.-T., Xing,
  D., and Sun, J.}
\newblock Superionic silica-water and silica-hydrogen compounds in the deep
  interiors of uranus and neptune.
\newblock {\em Phys. Rev. Lett. 128\/} (Jan 2022), 035702.

\bibitem{gayer2016freecad}
{\sc Gayer, D., O'Sullivan, C., and et.al.}
\newblock Freecad visualization of realistic 3d physical optics beams within a
  cad system-model.
\newblock In {\em Millimeter, Submillimeter, and Far-Infrared Detectors and
  Instrumentation for Astronomy VIII\/} (2016), vol.~9914, International
  Society for Optics and Photonics, p.~99142Y.

\bibitem{geuzaine2009gmsh}
{\sc Geuzaine, C., and Remacle, J.-F.}
\newblock Gmsh: A 3-d finite element mesh generator with built-in pre-and
  post-processing facilities.
\newblock {\em International journal for numerical methods in engineering 79},
  11 (2009), 1309--1331.

\bibitem{Hanfland1993}
{\sc Hanfland, M., Hemley, R.~J., and Mao, H.-k.}
\newblock Novel infrared vibron absorption in solid hydrogen at megabar
  pressures.
\newblock {\em Phys. Rev. Lett. 70\/} (Jun 1993), 3760--3763.

\bibitem{He-Nature2022}
{\sc He, Y., and et~al.}
\newblock Superionic iron alloys and their seismic velocities in earth’s
  inner core.
\newblock {\em Nature 602\/} (2022), 258--262.

\bibitem{Hemley1994}
{\sc Hemley, R.~J., Soos, Z.~G., Hanfland, M., , and Mao, H.-k.}
\newblock Charge-transfer states in dense hydrogen.
\newblock {\em Nature 369\/} (1994), 384--387.

\bibitem{Hou-Nature2021}
{\sc Hou, M., He, Y., Jang, B., Sun, S., Zhuang, Y., Deng, L., Tang, R., Chen,
  J., Ke, F., Meng, Y., Prakapenka, V., Chen, B., Shim, J., Liu, J., Kim, D.,
  Hu, Q., Pickard, C., R.J., N., and H-K., M.}
\newblock Superionic iron oxide–hydroxide in earth’s deep mantle.
\newblock {\em Nat. Geosci. 14\/} (2021), 174--178.

\bibitem{Kauzmann}
{\sc Kauzmann, W.}
\newblock {\em Quantum Chemistry}.
\newblock Academic, New York, 1957.

\bibitem{kuznetsov1991practice}
{\sc Kuznetsov, D.}
\newblock Practice of short-term prediction of earthquakes: astro-,
  cosmo-geophysical impulses of vernadsky-vlasov-vorobjev-prigozhin on the
  vertical sequence of underground electrodes at pedinstitute fault at the
  magnetic meridian of petropavlovsk-kamchatsky.
\newblock {\em All-Union Institute for science and technical information
  (VINITI), Moscow}, 3256-V91 (1991), 1--9.

\bibitem{larin1980hypothesis}
{\sc Larin, V.}
\newblock Hypothesis of a primordially hydride earth.
\newblock {\em Moscow, Izdatel'stvo Nedra\/} (1980).

\bibitem{Lee}
{\sc Lee, K.}
\newblock A mixed problem for hyperbolic equations with time-dependent domain.
\newblock {\em J. Math. Anal. Appl. 16\/} (1966), 471--495.

\bibitem{LoggMardalEtAl2012a}
{\sc Logg, A., Mardal, K.-A., Wells, G.~N., et~al.}
\newblock {\em Automated Solution of Differential Equations by the Finite
  Element Method}.
\newblock Springer, 2012.

\bibitem{Millot-Nature2018}
{\sc Millot, M., Hamel, S., Rygg, J., Celliers, P.~M., Collins, G.~W., Coppari,
  F., Fratanduono, D.~E., Jeanloz, R., Swift, D.~C., and Eggert, J.~H.}
\newblock Experimental evidence for superionic water ice using shock
  compression.
\newblock {\em Nat. Phys. 14\/} (2018), 297--302.

\bibitem{Nabighian}
{\sc Nabighian, M.~N.}
\newblock {\em Electromagnetic methods in applied geophysics}.
\newblock Tulsa, Okla: Soc. of Exploration Geophysicists, 1993.

\bibitem{Nishi-Nature2014}
{\sc Nishi, M., Irifune, T., Tsuchiya, J., Tange, Y., Nishihara, Y., Fujino,
  K., and Higo, Y.}
\newblock Stability of hydrous silicate at high pressures and water transport
  to the deep lower mantle.
\newblock {\em Nat. Geosci. 7\/} (2014), 224--227.

\bibitem{Varostos}
{\sc P~Varostos, P., and Alexopoulos, K.}
\newblock Physical properties of the variation of the electric fields of the
  earth preceding earthquakes.
\newblock {\em Tectonophysics 110\/} (1984), 73--98.

\bibitem{Pao}
{\sc Pao, C.}
\newblock {\em Nonlinear Parabolic and Elliptic Equations.}
\newblock Springer, 1992.

\bibitem{rudykh1989nonstationary}
{\sc Rudykh, G.~A., Sidorov, N.~A., and Sinitsyn, A.~V.}
\newblock Nonstationary solutions of the two-particle vlasov–maxwell system.
\newblock {\em Dokl. Math. 34}, 8 (1989), 700--701.

\bibitem{sinitsyn2011kinetic}
{\sc Sinitsyn, A., Vedenyapin, V., and Dulov, E.}
\newblock {\em Kinetic Boltzmann, Vlasov and related equations}.
\newblock Elsevier, 2011.

\bibitem{squillacote2007paraview}
{\sc Squillacote, A.~H., Ahrens, J., Law, C., Geveci, B., Moreland, K., and
  King, B.}
\newblock {\em The paraview guide}, vol.~366.
\newblock Kitware Clifton Park, NY, 2007.

\bibitem{Sugimura-Chem2012}
{\sc Sugimura, E., Komabayashi, T., Ohta, K., Hirose, K., Ohishi, Y., and
  Dubrovinsky, L.}
\newblock Experimental evidence of superionic conduction in $h2o$ ice.
\newblock {\em J. Chem. Phys. 137\/} (2012), 194505.

\bibitem{Surkov}
{\sc Surkov, V., and Hayakawa, M.}
\newblock {\em Ultra and Extremely Low Frequency Electromagnetic Fields}.
\newblock Tokyo: Springer Japan, 2014.

\bibitem{Uyeda}
{\sc Uyeda, S., Nagao, T., Orihara, Y., Yamaguchi, T., and Takahashi, I.}
\newblock Geoelectric potential changes: possible precursors to earthquakes in
  japan.
\newblock {\em Proc Natl Acad Sci USA 97}, 9 (2000), 4561--4566.

\bibitem{vernadsky1912gas}
{\sc Vernadsky, V.}
\newblock Gas exchange in earth's crust.
\newblock {\em Proceedings of St. Petersburg Royal Academy of Sciences 6\/}
  (1912), 141--162.

\bibitem{vlasov1961many}
{\sc Vlasov, A.~A.}
\newblock {\em Many-particle theory and its application to plasma.}
\newblock Gordon and Breach, 1961.

\end{thebibliography}
\bibliographystyle{acm}

\end{document}